\documentclass[conference]{IEEEtran}
\usepackage{cite}
\usepackage{amsmath,amssymb,amsfonts}
\usepackage{algorithmic}
\usepackage{graphicx}
\usepackage{graphics}
\usepackage{textcomp}
\usepackage{fancyhdr}
\usepackage[hyphens]{url}

\usepackage{subcaption}
\usepackage{multirow}
\usepackage[inline]{enumitem}
\usepackage[table,x11names]{xcolor}

\usepackage{authblk}

\def\BibTeX{{\rm B\kern-.05em{\sc i\kern-.025em b}\kern-.08em
    T\kern-.1667em\lower.7ex\hbox{E}\kern-.125emX}}

\definecolor{yellow}{rgb}{1,1,0}

\fancypagestyle{firstpage}{
  \fancyhf{}

  \fancyhead[C]{\normalsize{Submission Under Review for an IEEE Conference}} 
  \fancyfoot[C]{\thepage}
}

\title{Direct CMOS Implementation of
Neuromorphic Temporal Neural Networks for Sensory Processing\vspace{-0.5ex}}

\author{Harideep Nair\vspace{-1.5ex}}
\author{John Paul Shen}
\author{James E. Smith}
\affil{Carnegie Mellon University\vspace{-1.5ex}}

\begin{document}
\maketitle
\thispagestyle{firstpage}
\pagestyle{plain}


\begin{abstract}

Temporal Neural Networks (TNNs) use time as a resource to represent and process information, mimicking the behavior of the mammalian neocortex. This work focuses on implementing TNNs using off-the-shelf digital CMOS technology. A microarchitecture framework is introduced with a hierarchy of building blocks including: multi-neuron \textit{columns}, multi-column \textit{layers}, and multi-layer TNNs.
We present the \textit{direct} CMOS gate-level implementation of the multi-neuron column model as the key building block for TNNs. Post-synthesis results are obtained using Synopsys tools and the 45 nm CMOS standard cell library.
The TNN microarchitecture framework is embodied in a set of characteristic equations for assessing the total gate count, die area, compute time, and power consumption for any TNN design. 
We develop a multi-layer TNN prototype of 32M gates. In 7 nm CMOS process, it consumes only 1.54 mm\textsuperscript{2} die area and 7.26 mW power and can process 28x28 images at 107M FPS (9.34 ns per image). We evaluate the prototype's performance and complexity relative to a recent state-of-the-art TNN model.


\end{abstract}

\section{Introduction}

\subsection{Temporal Neural Networks}

Temporal Neural Networks (TNNs) \cite{smith2017space,smith2018space} are a class of spiking neural networks (SNNs) that encode and process information in temporal form using precise \textit{spike timings} to represent information \cite{vanrullen2005spike} unlike other forms of SNNs that use \textit{spike rates} for information encoding and processing. Furthermore, unlike artificial neural networks (ANNs), TNNs strive for biological plausibility with the goal of achieving brain-like capability and brain-like efficiency.

\begin{figure}[t]
    \centering
    \includegraphics[width=3.5in]{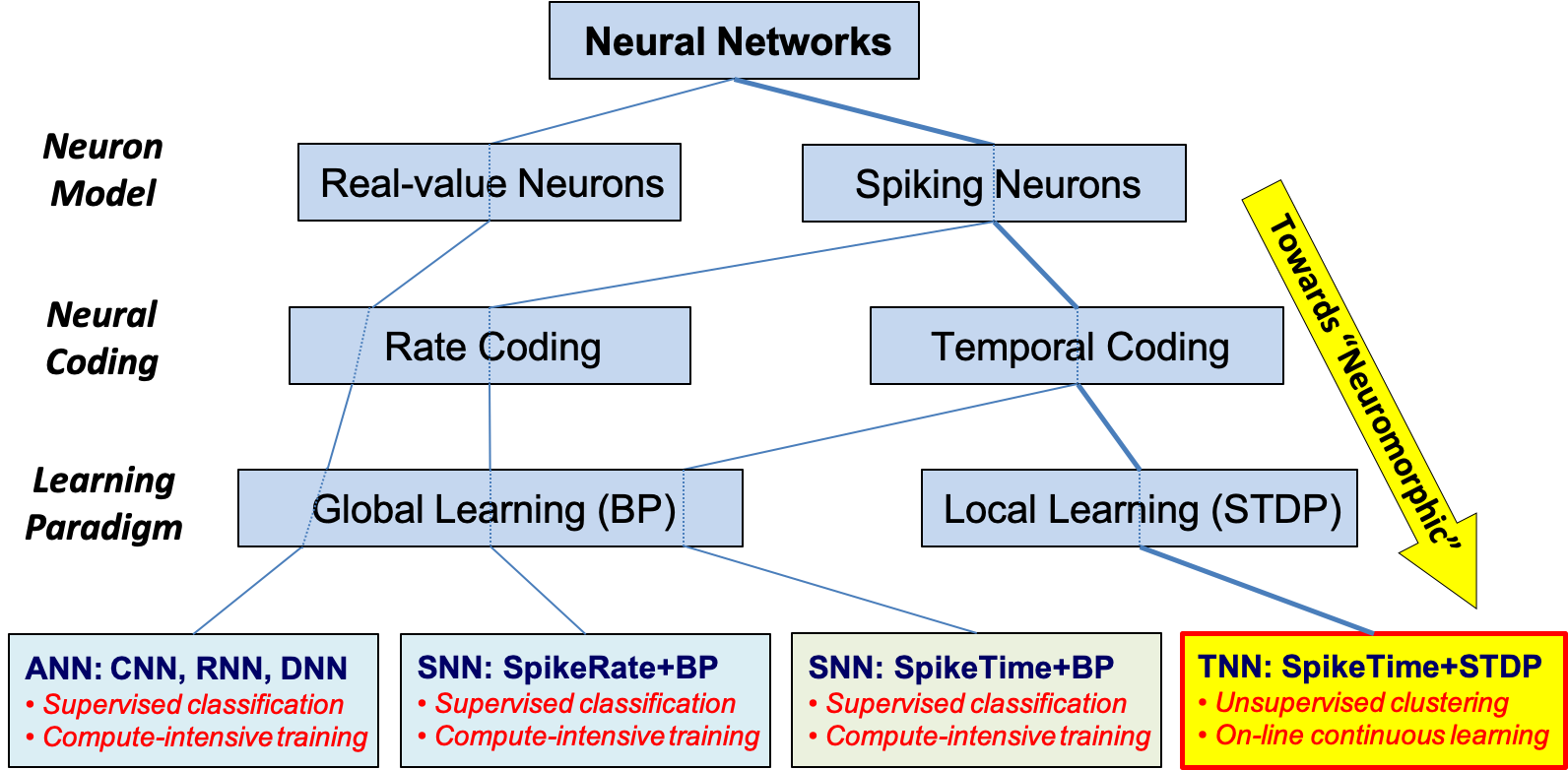}
    \caption{Neural Network Taxonomy}
    \label{tax}
\end{figure}

TNNs are composed of a hierarchy of neurons that perform temporal functions on temporally encoded values. TNN neurons communicate via spikes, encode information using relative spike timings, and employ STDP local learning, as shown in Figure \ref{tax}.  TNNs
are trained via Spike Timing Dependent Plasticity (STDP) \cite{guyonneau2005neurons}. With STDP, a synapse’s weight is updated based only on the relative timings of its incoming spike (from a pre-synaptic neuron) and the outgoing spike, if any, (from its post-synaptic neuron). Hence, STDP training is localized and has the potential for enabling unsupervised, online, and continuous learning, without requiring compute-intensive back propagation commonly employed in ANNs and SNNs. 
These distinctive attributes of TNNs adhere strongly to biological plausibility and make them truly neuromorphic.

\subsection{Our Approach}

TNN research has a long history. Some of the historical contributions are highlighted here.
In 1989, Thorpe and Imbert \cite{thorpe1989biological, thorpe1990spike} made a persuasive experimentally-supported argument for inter-neuron communication via precisely timed spikes. 
Hebb \cite{hebb1949organization} observed that repeated temporal coincidence between a synapse’s input spike and its post-synaptic neuron’s output spike tends to increase the synapse’s weight. 
In 1983, Levy and Steward \cite{levy1983temporal} established the classic STDP update rules. 
Gerstner et al. \cite{gerstner1996neuronal} proposed a theoretical foundation for STDP as an integral part of a computing paradigm. 

This work builds on recent works in \cite{smith2017space,smith2018space}, \cite{smith2019newtonian} which laid the foundation of TNNs as space-time computing networks based on a rigorous space-time algebra. The author in \cite{smith2019fcrc, smith2018tut} suggested building a \textit{silicon neocortex} 
by examining the hierarchical organization of biological neural networks to formulate an analogous hierarchical organization for TNNs. As computer architects, we follow this proposed approach and focus on direct hardware implementation of TNNs.

\subsection{Our Contributions}

\begin{figure}[t]
    \centering
    \includegraphics[width=3.0in]{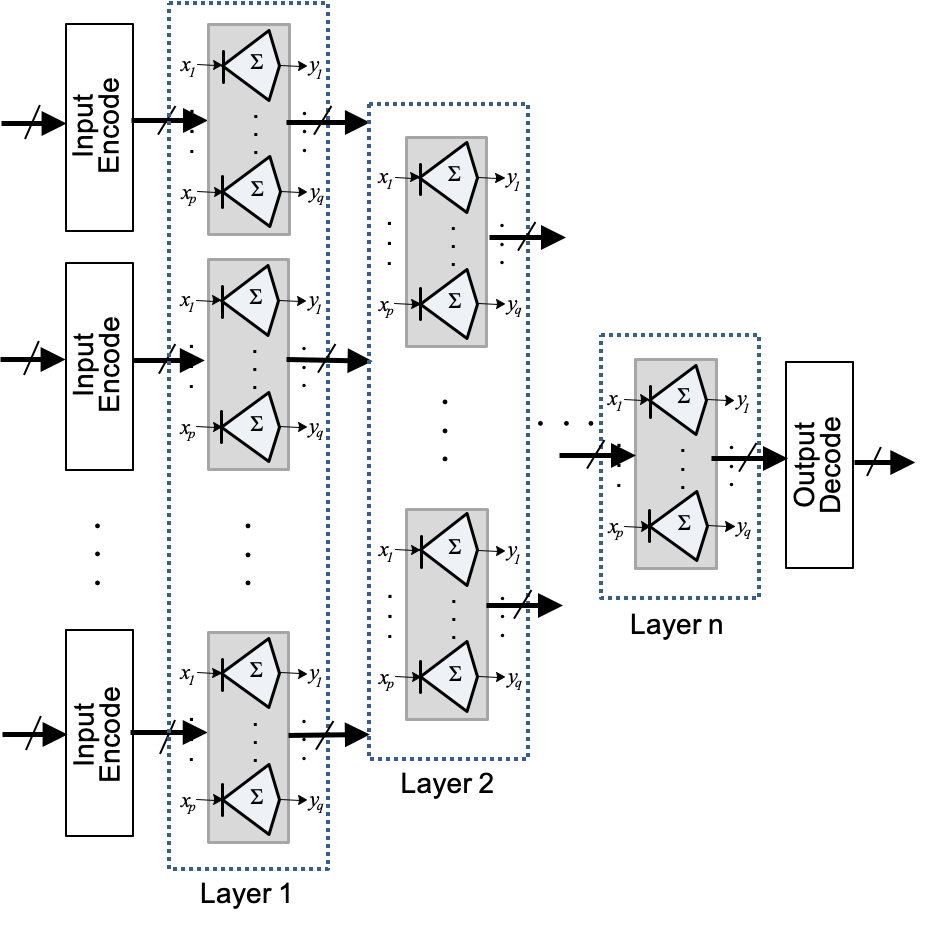}
    \caption{Generic Temporal Neural Network Organization: with stacking of multi-neuron columns in each layer and cascading of multi-column layers into a multi-layer TNN, with dimensions: \textbf{TNN\{[s\textsubscript{1}x(p\textsubscript{1}xq\textsubscript{1})] + [s\textsubscript{2}x(p\textsubscript{2}xq\textsubscript{2})] + ... + s\textsubscript{n}x(p\textsubscript{n}xq\textsubscript{n})]\}} where s\textsubscript{i} denotes the number of (p\textsubscript{i}xq\textsubscript{i}) columns in layer i. TNN is bookended by input-encode and output-decode layers.
    }
    \label{fig2_new}
\end{figure}

In this work, we explore the practical feasibility of direct hardware implementation of TNNs using off-the-shelf digital CMOS technology and design tools. In a direct CMOS implementation, the actual hardware clock cycle is used as the basic time unit for temporal processing. We focus on defining a TNN microarchitecture and implementing its hierarchy of building blocks.
From the bottom up, these building blocks include: \textit{neurons}, multi-neuron \textit{columns}, and multi-column \textit{layers}, resulting in multi-layer TNNs, as shown in Figure  \ref{fig2_new}.

We present gate-level  designs (in Verilog) of neurons and columns, including gate-level implementation of STDP (unsupervised) and R-STDP (supervised) learning rules. We obtain post-synthesis results using Synopsys tools. These designs are evaluated based on the metrics of: gate count, die area, critical path delay/compute time, and power consumption. 

Key contributions of this work include:
%
%
%
\begin{itemize}[itemsep=1pt, topsep=1pt]
    \item A microarchitecture framework for direct CMOS implementation of TNNs, based on building blocks of neurons, columns of neurons, and layers of columns.
    \item Gate-level designs of: 1) a scalable neuron with ramp-no-leak (RNL) response function and STDP and R-STDP learning rules, and 2) a scalable column with winner-take-all (WTA) lateral inhibition.
    \item Characterizing equations for total gate counts when scaling the synapse count \textit{p} for a neuron, and the neuron count \textit{q} for a column (useful for exploring design space).
    \item Post-synthesis evaluation of Area (mm\textsuperscript{2}), Delay/Time (ns), and Power (mW) metrics for the neuron and column designs. Our logic synthesis is based on the standard cell library for the 45 nm CMOS process node.
    \item A prototype 2-layer TNN design (32M gates), with one unsupervised STDP layer and one supervised R-STDP layer, is used  to demonstrate the potential of future TNN-based online sensory processing units. We show fast training convergence with good accuracy on MNIST.
    \item Initial results show that this prototype TNN, when scaled to 7 nm CMOS, can process a 28x28 MNIST image in 9.34 ns while consuming only 7.26 mW and requiring 
    only 1.54 mm\textsuperscript{2} die area.
\end{itemize}

\section{Background and Prior Research}

\subsection{Deep Artificial Neural Networks}
Deep Artificial Neural Networks (ANNs), including convolutional neural networks (CNNs) and recurrent neural networks (RNNs), are dominant paradigms for performing human-like sensory processing using machine learning (ML) techniques. These networks effectively support ML applications such as image classification and speech recognition.

Deep ANNs typically employ: 1) tensor processing with high-dimensional matrix multiplications, and 2) supervised global learning with back propagation and stochastic gradient descent. Even though highly effective, neither of these techniques are biologically plausible and require an enormous amount of linear algebraic computation. Since 2012, the computation required for training deep ANNs has been doubling every 3.4 months, or increasing at the rate of 10x/year \cite{openai}.
%

TNNs are fundamentally different and are based on a computing paradigm that does not involve high dimensional tensor processing. TNNs perform computation based on spike timing relationships and do not use linear algebraic operations. The learning paradigm for TNNs is localized and predominantly unsupervised, and is capable of online and continuous learning. In TNNs, inference and training can occur simultaneously.

\subsection{Spiking Neural Networks}
Spiking Neural Networks (SNNs) use spikes to communicate between neurons and are a broader class of neural networks that includes TNNs as shown in Figure \ref{tax}. However, many SNNs in the literature employ rate encoding as opposed to TNNs which use temporal encoding. In rate encoding, \textit{spike rates} are used to represent values whereas in temporal encoding, relative \textit{spike times} are used.

Many SNNs also use backpropagation for learning, similar to  deep ANNs. Multilayer SNNs proposed in \cite{o2016deep, lee2016training, neftci2017stochastic} employ both rate encoding and backpropagation, whereas SNNs in \cite{diehl2015unsupervised, tavanaei2017multi,
tavanaei2017spiking, querlioz2013immunity, brader2007learning} use rate encoding coupled with STDP. The works in \cite{liu2017mt, mostafa2017supervised} implement feedforward SNNs with temporal encoding but use backpropagation for learning. TNNs use spike times for temporal encoding and STDP for local learning.

\subsection{Neuromorphic Hardware}
In recent years, several experimental neuromorphic chips, implemented in standard digital CMOS, have been introduced. 

TrueNorth is a digital neuromorphic chip introduced by IBM in 2014 \cite{merolla2014million} fabricated using 28 nm CMOS technology. It consists of one million neurons distributed across 4096 cores, with each core containing 256 leaky-integrate-and-fire (LIF) neurons, each with 256 synapses. Synaptic weights can be set to either 0 or 1 (1 bit). The chip runs at 1 KHz with 4-bit resolution for time steps (0 to 15). Time delays are stored and processed as 4-bit binary values, making this an \textit{indirect} implementation. Spikes are communicated as packets using address event representation (AER). The authors use offline training for their performance demonstration. 

Intel unveiled Loihi in 2018 as their flagship neuromorphic chip built in 14 nm process \cite{davies2018loihi}. Each chip consists of 130,000 LIF neurons distributed across 128 cores; each core consists of 1024 neurons, each with 1024 synapses. It can support 1-9 bit synaptic weights. Spikes are represented as packetized messages. Intel has provided Loihi boards for cloud access and experimentation by researchers.

ODIN \cite{frenkel20180} is a neuromorphic ASIC designed in 28 nm CMOS, consisting of 256 neurons with 256 synapses each. It implements a version of STDP, Spike Dependent Synaptic Plasticity (SDSP), for local learning. It implements 3-bit synaptic weights. Spikes are represented as AER packets.

FlexLearn \cite{baek2019flexlearn} is a recent neuromorphic processor designed in 45 nm CMOS. The 128-core FlexLearn processor requires 410 mm\textsuperscript{2} of die area and consumes 13,981 mW. 
FlexLearn provides more flexibility in supporting complex learning rules.

These neuromorphic chips provide highly configurable fabric for experimentation. Our focus is on developing a TNN architecture and the microarchitecture of its key building blocks. Our eventual goal is creating a framework for the design of purpose-specific TNNs for neuromorphic edge-native online sensory processing  that require only a few mm\textsuperscript{2} die area and consume only a few mW of power. 

\section{TNN Organization and Operation}

We leverage insights from both conventional computer systems as well as biological neocortex systems. Both are hierarchical systems with multiple levels of abstraction.
The building blocks in our hierarchical TNN microarchitecture mimic those of biological neural networks (Figure \ref{neuro_abst}). TNN building blocks include \textit{neurons}, multi-neuron \textit{columns}, and multi-column \textit{layers}. This paper focuses on columns as the primary building blocks for multi-layer TNNs (Figure \ref{fig2_new}). 
\begin{figure}[t]
    \centering
    \begin{subfigure}[b]{0.225\textwidth}
        \centering
        \includegraphics[width=\textwidth]{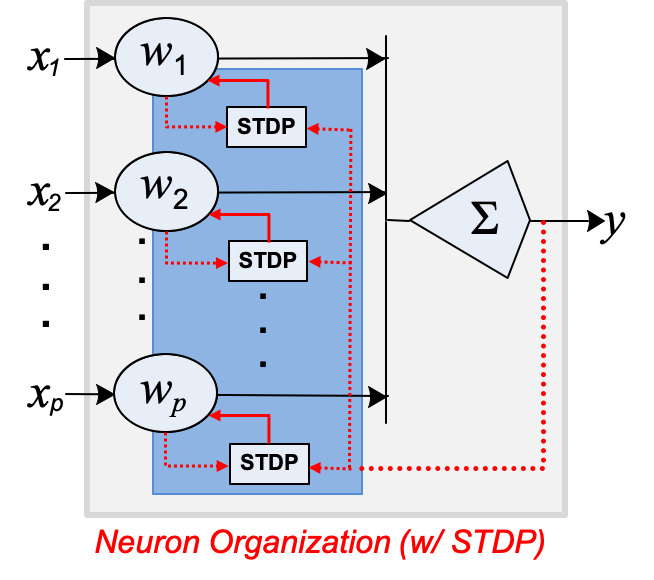}
        \caption{Neuron: \textit{p} Synapses, STDP}
        \label{fig3a_new}
    \end{subfigure}%
    \begin{subfigure}[b]{0.295\textwidth}
        \centering
        \includegraphics[width=\textwidth]{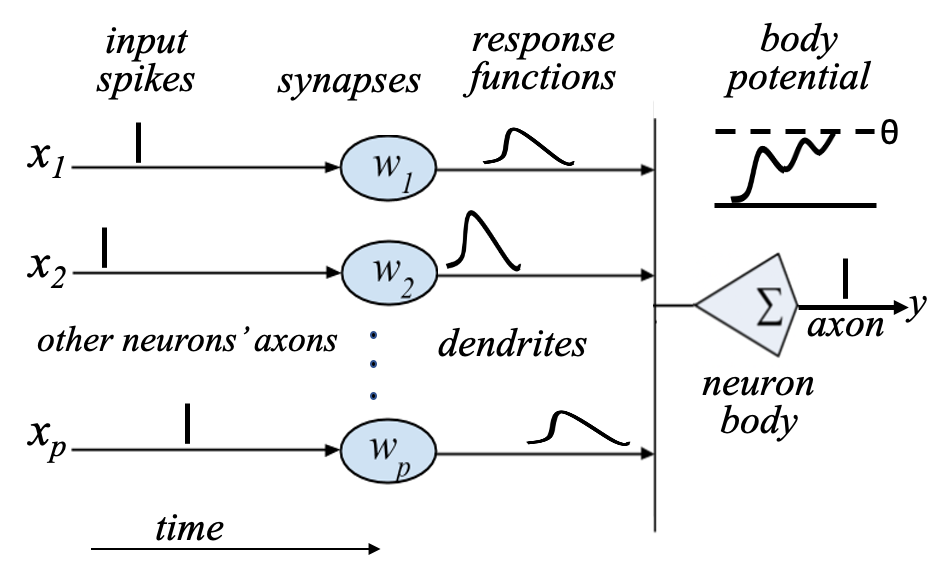}
        \caption{Neuron Operation}
        \label{fig3b_new}
    \end{subfigure}
    \caption{A TNN Neuron Model}
\end{figure}
\begin{figure}[t]
    \centering
    \begin{subfigure}[b]{0.225\textwidth}
        \centering
        \includegraphics[width=\textwidth]{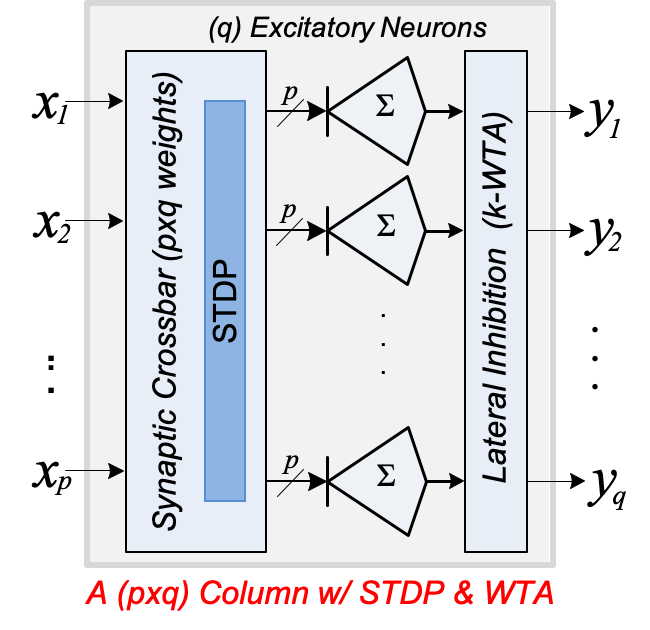}
        \caption{Column: q Neurons \& WTA}
        \label{fig4a_new}
    \end{subfigure}%
    \begin{subfigure}[b]{0.29\textwidth}
        \centering
        \includegraphics[width=\textwidth]{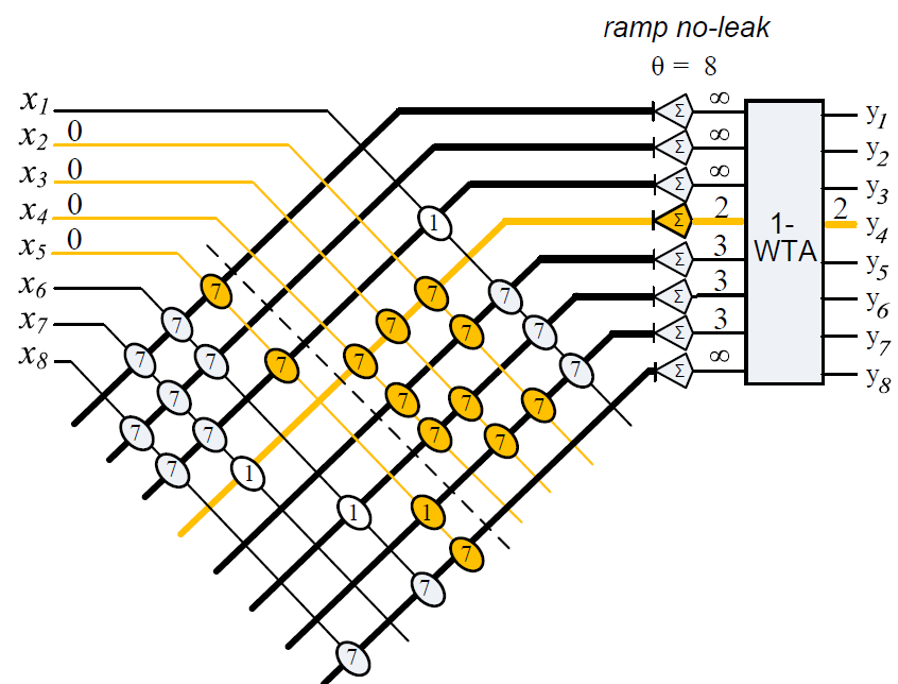}
        \caption{Column Operation}
        \label{fig4b_new}
    \end{subfigure}
    \caption{A TNN Column Model}
\end{figure}
\subsection{Key TNN Building Blocks}
%

As shown in Figure \ref{fig3a_new}, each neuron has \textit{p} synaptic inputs and one output. Each synaptic input carries a synaptic weight, which is updated locally based on the relative timing of the incoming spike to that synapse and the outgoing spike from the neuron body. The rules for updating synaptic weights constitute the STDP learning method. 

As shown in Figure \ref{fig4a_new}, a column is a stacking of \textit{q} parallel neurons. Every neuron in a column shares the same set of \textit{p} inputs. There is a \textit{pxq} synaptic crossbar containing \textit{pxq} synaptic weights, each of which is independently updated by STDP. On the output side of the \textit{q} neurons, \textit{lateral inhibition} is performed by selecting the earliest spiking \textit{k} neurons from among the \textit{q} neurons as the winners (\textit{k}-WTA). Output spiking is disabled for non-winning neurons. Typically \textit{k}=1 is used.

\begin{figure}[t]
    \centering
    \begin{subfigure}[b]{0.212\textwidth}
        \centering
        \includegraphics[width=\textwidth]{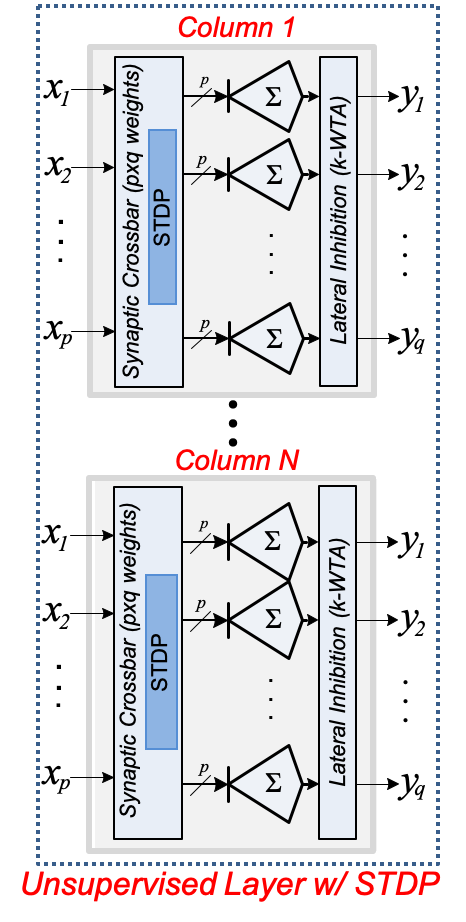}
        \caption{Unsupervised Layer}
        \label{fig5a_new}
    \end{subfigure}%
    \begin{subfigure}[b]{0.212\textwidth}
        \centering
        \includegraphics[width=\textwidth]{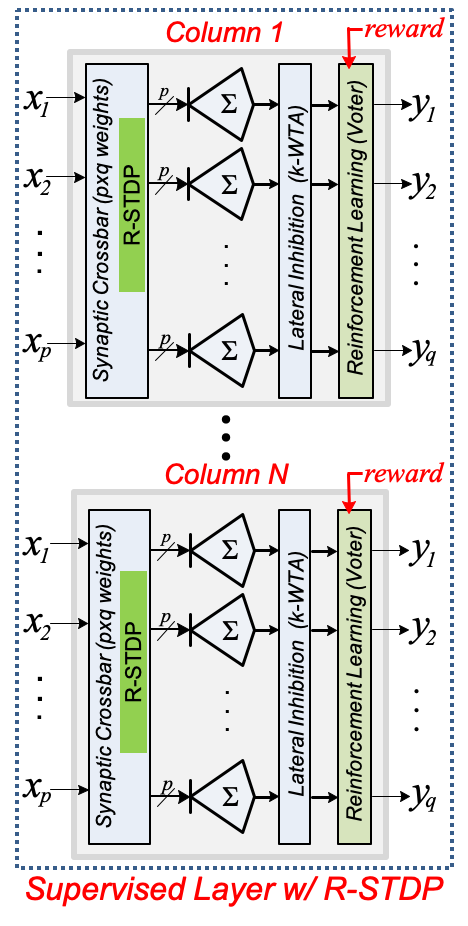}
        \caption{Supervised Layer}
        \label{fig5b_new}
    \end{subfigure}
    \caption{TNN Multi-Column Layers: Two Types}
    \label{fig5_new}
\end{figure}

This paper presents the CMOS implementation of the neuron (Section \ref{neuron_sec}) and the column (Section \ref{column_sec}). In Section \ref{stdp_sec}, STDP rules for updating synaptic weights are given. The baseline STDP method is unsupervised. We also introduce a variation, called \textit{reinforcement} STDP, which is similar to the \textit{reward modulated} STDP in \cite{mozafari2019bio}. R-STDP is usually deployed in the last layer of a multi-layer TNN. Figure \ref{fig5_new} illustrates the two types of TNN multi-column layers: Unsupervised Layer with STDP and Supervised Layer with R-STDP.
\subsection{Temporal Encoding and Processing}
A distinctive attribute of TNNs involves the use of temporal encoding. With temporal encoding, information is represented by relative timings of spikes. In a TNN, computation occurs in volleys or waves of spikes. A volley consists of at most one spike per synaptic input (some will have no spike). The value represented by each input spike is based on its spike time relative to the first spike in the volley. The first spike in the volley represents a value of 0 and subsequent spikes are assigned increasing values based on increasing delays relative to the first spike. 

\begin{figure}[t]
    \centering
    \includegraphics[width=3.35in]{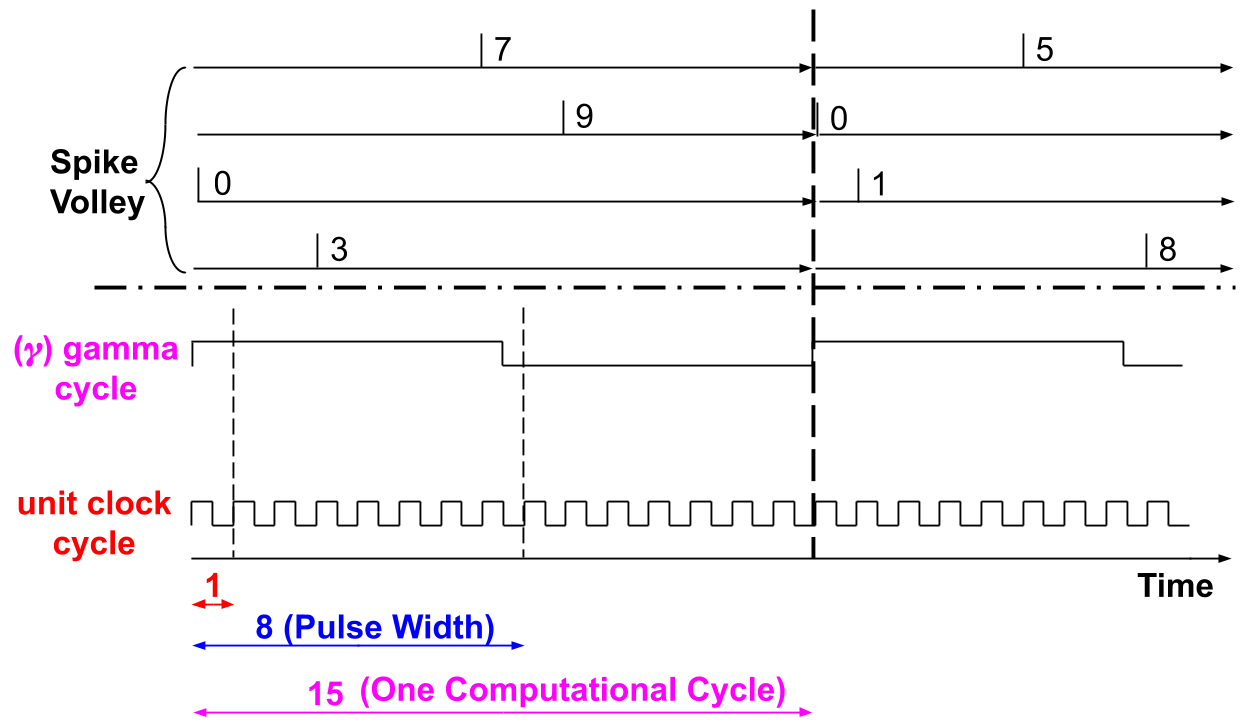}
    \caption{Temporal Encoding and Processing}
    \label{tcoding}
\end{figure}

Temporal encoding is illustrated in 
Figure \ref{tcoding}. 
If no spike occurs on an input, it is assigned the symbol ``$\infty$''.
It has been proposed, with experimental support, that computational volleys are synchronized by inhibitory oscillations at gamma frequencies (50-100 Hz) \cite{gray1989oscillatory}. These gamma oscillations divide neocortical processing into alternating inhibitory and excitatory phases. The inhibitory part of the cycle essentially performs a reset function and the excitatory part performs computation. This leads to a temporal coding interval of 5-10 msec during which spikes are transmitted in a coordinated volley.
Neuron spiking behavior has been shown to be repeatable to within 1 msec \cite{butts2007temporal, mainen1995reliability}. A 5-10 msec window combined with the 1 msec encoding precision implies only 3-4 bits of precision is needed within the encoding window. Consequently, the computing model used in this paper is based on low resolution integers.

In this work, temporal encoding and processing are employed with the actual hardware clock cycle directly serving as the basic time unit. This work uses 3 bits of precision for encoding and synaptic weights.
Spikes in a volley are represented using pulses which are a form of unary encoding and volleys are separated using gamma clock cycles. With unary encoding, it takes up to 7 time units to encode a 3-bit value. To allow additional time for a column to process a spike volley, the gamma cycle is extended to 15 time units. This is explained in further detail in Section \ref{fsmsyn_sec}.

In summary, the proposed design uses two clocks. The \textit{unit time clock} is the finest temporal resolution in the computation model and is also the synchronizing clock used in the digital hardware. The \textit{gamma clock} frames the computing window and is the time required for a column to communicate and process spike volleys and update synaptic weights.

\section{Neuron Implementation}
\label{neuron_sec}
This work focuses on the SRM0 excitatory neuron model based on the widely-used Spike Response Model \cite{kistler1997reduction} shown in Figure \ref{fig3b_new}. This section presents the components of this model and their detailed gate level designs along with corresponding gate count equations (using equivalent gate counts in terms of 4-input AND gates).
For gate level designs and analysis, we set the maximum weight value $w_{max}$ = 7.

\subsection{Synaptic Response Functions}
As shown in Figure \ref{fig3b_new}, a \textit{synapse} connects the \textit{axon} (output) of a pre-synaptic neuron and a \textit{dendrite} (input) of the post-synaptic neuron. An SRM0 neuron takes multiple input spikes and generates a response function for each spike based on its corresponding synaptic weight. All the individual response functions are then integrated to form the neuron’s membrane potential. When (and if) the membrane potential crosses a threshold, the neuron fires an output spike on its axon.  

\begin{figure}[t]
        \centering
        \includegraphics[width=3.3in]{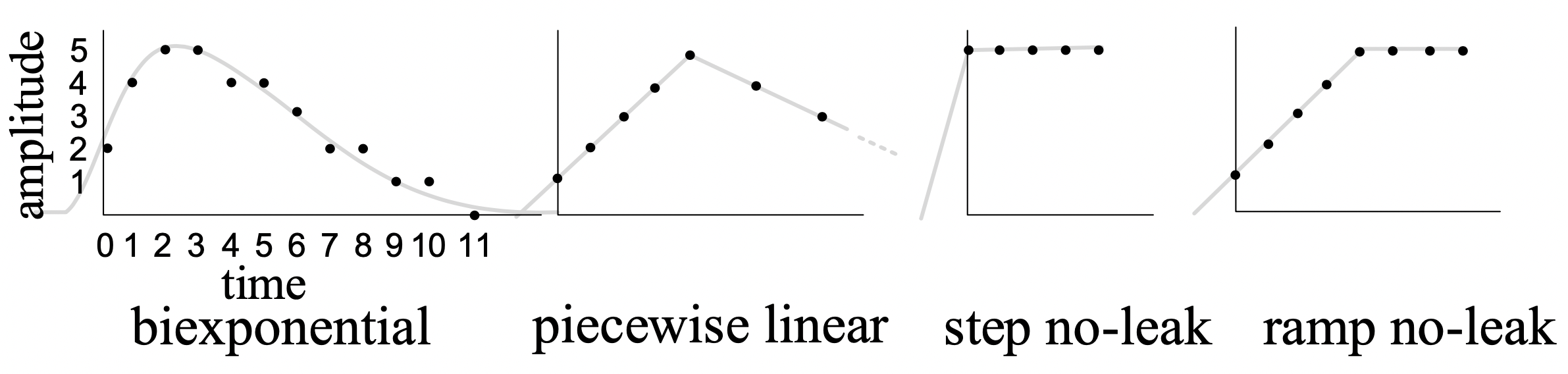}
        \caption{Four Well-known  Discretized Response Functions}
        \label{respfunc}
\end{figure}

Although a wide variety of response functions may be used as shown in Figure \ref{respfunc}, the response function of interest here is the ramp-no-leak (RNL) function due to its temporal computational benefits and implementation efficiency. The RNL function increases by a unit step at every time unit until it reaches its peak and then remains constant until it is reset prior to the next computation cycle. The ``ramp'' allows responses from different synapses to be distributed temporally based on the synaptic strengths (weights), which proves to be particularly powerful for TNNs that operate temporally. Note that this model doesn't ``leak''. This is based on arguments that the leak is actually just a reset mechanism \cite{guyonneau2005neurons, masquelier2007unsupervised}.

\subsection{FSM: Synapse Modeling}
\label{fsmsyn_sec}
\begin{figure}[t]
        \centering
        \includegraphics[width=3.3in]{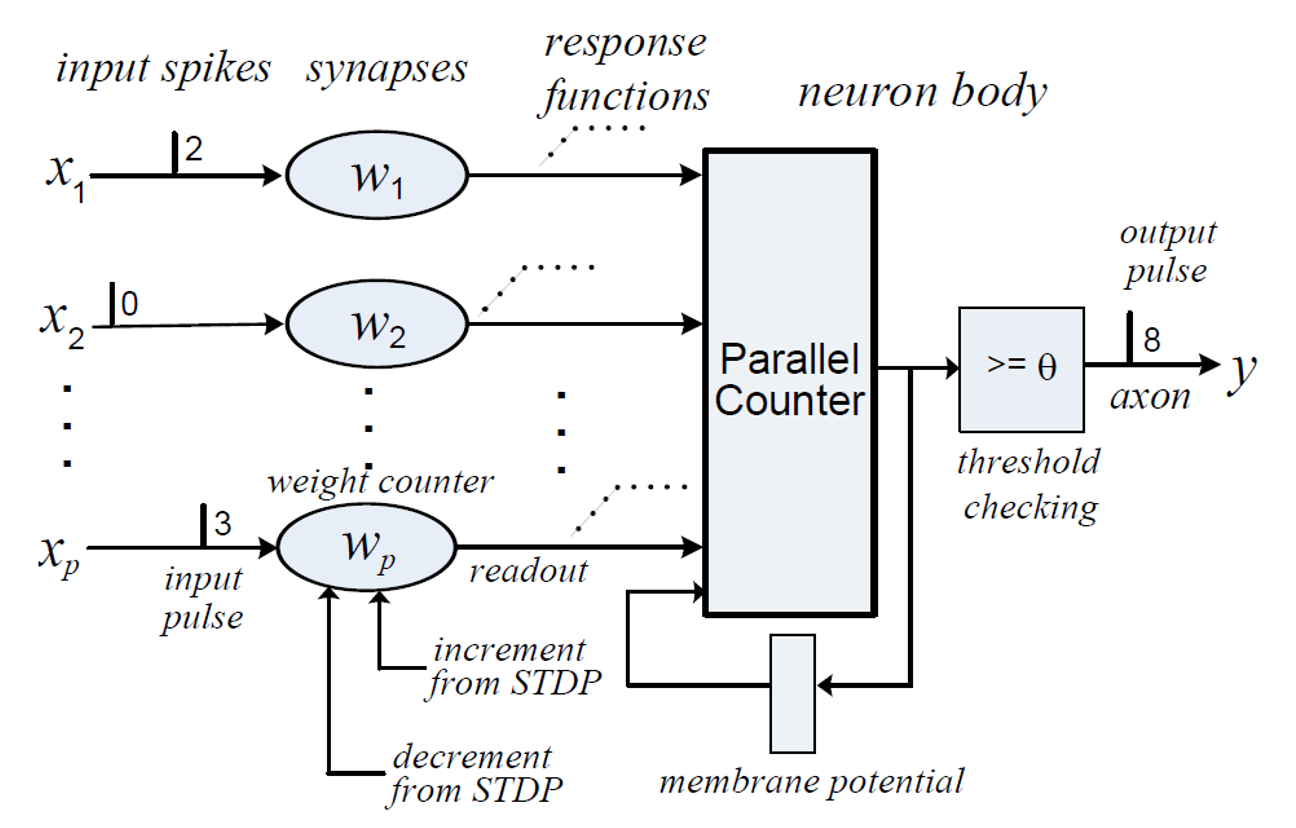}
        \caption{SRM0 Neuron Implementing RNL Response Function}
        \label{pt}
\end{figure}
Figure \ref{pt} shows the block diagram for the proposed SRM0 neuron implementing ramp-no-leak response function. Its operation consists of three main stages: 1) temporal arrival of input spikes, 2) serial thermometer readout of RNL response functions based on the corresponding synaptic weights, and 3) binary accumulation of thermometer-coded response functions into the membrane potential.
%
Synaptic weights are implemented as binary counters. If the maximum weight is $w_{max}$, the number of counter bits is $ceiling(log_2(w_{max} + 1))$. The counter has three modes, two controlled by STDP (described in Section \ref{stdp_sec}): increment (up to $w_{max}$) and decrement (down to 0). The third \textit{readout} mode is controlled by the input pulse.
Readout mechanism is meticulously integrated into the same FSM used for storing synaptic weight and is described below.

As will become apparent, the synapses are the dominant hardware cost, so the synapse design must focus on minimizing hardware. A clever idea involves using a pulse width equal to $w_{max}+1$. The input pulse directly controls the counter readout. When the leading edge of an input pulse occurs (0 to 1 transition), the weight is read out and decremented until it reaches 0. An output of 1 is emitted each unit clock cycle until the counter reaches 0. This essentially converts the binary weight value in the counter to a serial thermometer code. 
After the counter reaches 0, it wraps around to $w_{max}$ and continues to count down until the trailing edge of the input pulse (1 to 0 transition) when
the weight in the counter is restored to its original value. Thus, once an input spike arrives, readout takes an additional 7 cycles. (Although we assume $w_{max}$ = 7 in this paper, this technique can be generalized to any $w_{max}$.)
During training, STDP (explained in Section \ref{stdp_sec}) takes another cycle. These coupled with 7 cycles for encoding give rise to a gamma period of 15 clock cycles.

In summary, a synapse and its weight are implemented with a counter FSM that can 1) increment, saturating at $w_{max}$, 2) decrement, saturating at 0, and 3) wrap-around decrementing, emitting an output of 1 prior to wrapping around and a 0 thereafter. Based on our logic design, not including STDP, each synapse requires a total of 61 equivalent gates (counting latches as 2 gates, FF as 5 gates) or $61p$ gates for \textit{p} synapses, most of which arise from the state transition logic.

\subsection{Neuron Body}
\begin{figure}[t]
        \centering
        \includegraphics[width=3.5in]{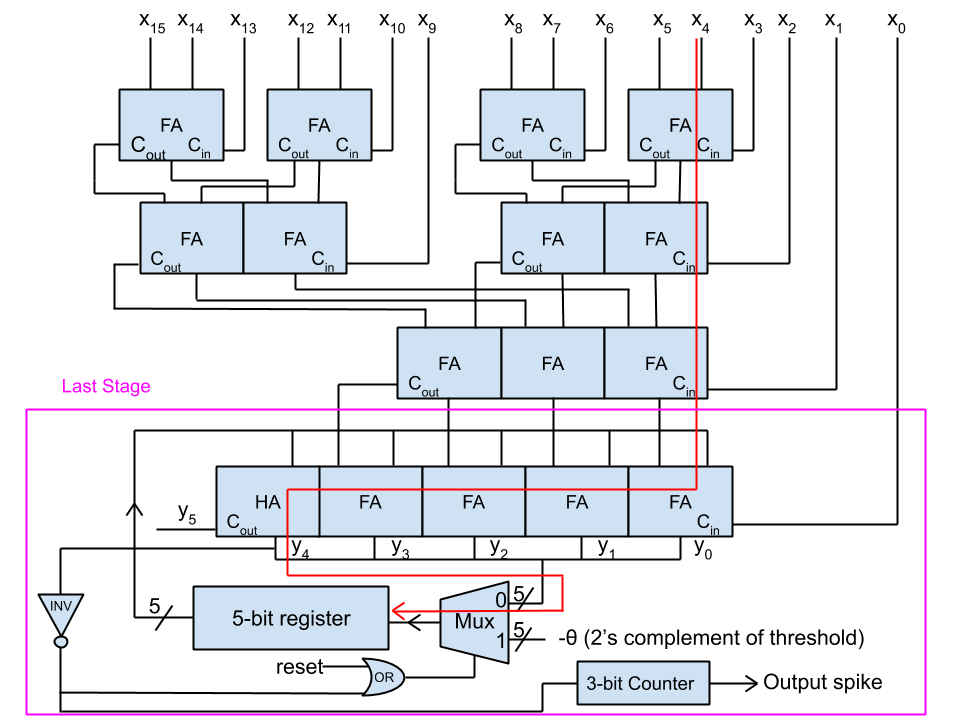}
        \caption{Neuron Body Potential Computation: a 16-input accumulator that stores into a 5-bit register and compares the stored value with a threshold to generate an output spike.}
        \label{acc}
\end{figure}
The neuron body is implemented as a parallel counter that adds the thermometer coded weights coming from the synapses, cycle by cycle, thereby accumulating the membrane potential as a sum of RNL response functions. When (and if) the parallel counter output reaches the threshold $\theta$, an output spike is emitted during that cycle.

Using the work of Parhami \cite{parhami1995accumulative}, the membrane potential accumulator can be implemented using ripple carry adders as fundamental units by integrating a (\textit{p}-1)-input parallel combinational counter and a ($log_2p+1$)-bit adder into one design. Figure \ref{acc} shows the logic diagram for a 16-input accumulator, with integrated output spike generation. For a \textit{p}-input accumulator, \textit{p}-1 inputs are accumulated into a ($log_2p$)-bit output, which is then added to the previous stored ($log_2p+1$)-bit value from the register with the one remaining input bit acting as carry-in. Note that the hierarchical configuration in Figure \ref{acc} allows all adder inputs to be efficiently utilized and is particularly optimal when \textit{p} is a power of 2. 

Furthermore, the accumulating register is initialized with (signed 2’s complement) -$\theta$ at every gamma cycle, which eliminates the need for any comparator for output spike generation. The  ($log_2p+1$)\textsuperscript{th} bit of the output can be used to determine if the accumulated body potential has crossed the threshold and trigger a 3-bit counter that generates an 8-unit time pulse (output spike). For a neuron body with \textit{p} synaptic inputs, the gate count comes out to be $5p+ 8log_2p+31$.
Conventional digital logic is used for implementing temporal encoding and processing with very efficient hardware. Small counters (FSMs) store and update synaptic weights, rather than much longer shift registers with unary representation of weights. A small adder tree sums multiple single-bit inputs to realize the summing of higher precision values over time. STDP and R-STDP designs are presented next.

\section{STDP \& R-STDP Implementation}
\label{stdp_sec}
STDP is a distinctive feature of TNNs. STDP learning is unsupervised and local to each synapse. It can perform inference and continuous learning at the same time. In this work, we  propose an STDP design that is both effective in learning and implementable using standard CMOS technology. 



\subsection{Proposed STDP Update Rules}
\begin{figure}
        \centering
        \includegraphics[width=3.2in]{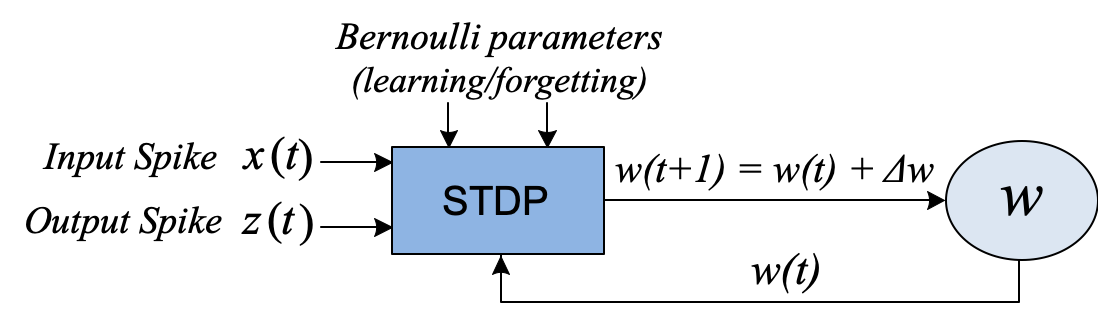}
        \caption{Spike Timing Dependent Plasticity (STDP) is implemented at each synapse. The synaptic weight is updated based on the synapse’s input spike time, its associated neuron’s output spike time and the current weight.}
        \label{stdp_illustrate}
\end{figure}

Our learning method is a customized version of the
classic Spike Timing Dependent Plasticity (STDP). STDP is implemented locally at each synapse as shown in Figure \ref{stdp_illustrate}. The proposed STDP learning rules are summarized in Table \ref{stdp}. Here, x(t) and z(t) represent input and output spiketimes respectively. $\Delta$w denotes change in weight and B($\mu$) represents a Bernoulli random variable with probability $\mu$.
\begin{table}[t]
  \centering
  \small
  \scalebox{0.84}{
  \begin{tabular}{|c|c||c|}
    \hline
    \multicolumn{2}{|c||}{\textbf{Input Conditions}} & \textbf{Weight Update}\\
    \hline
    \hline
    $x(t)\neq \infty$; & $x(t)\leq z(t)$ & $\Delta w=+B(\mu_{capture})*max(F(w),B(\mu_{min}))$\\
    \cline{2-3}
    $z(t)\neq \infty$ & $x(t)>z(t)$ & $\Delta w=-B(\mu_{backoff})*max(F(w),B(\mu_{min}))$\\
    \hline
    \multicolumn{2}{|c||}{$x(t)\neq \infty$; $z(t)=\infty$} & $\Delta w=+B(\mu_{search})$\\
    \hline
    \multicolumn{2}{|c||}{$x(t)=\infty$; $z(t)\neq \infty$} & $\Delta w=-B(\mu_{backoff})*max(F(w),B(\mu_{min}))$\\
    \hline
    \multicolumn{2}{|c||}{$x(t)=\infty$; $z(t)=\infty$} & $\Delta w=0$\\
    \hline
  \end{tabular}
  }
  \caption{Proposed STDP Update Rules}
  \label{stdp}
\end{table}

STDP update rules are divided into four major cases, corresponding to the four combinations of input and output spikes (represented by x(t) and z(t) respectively) being present ($\neq \infty$) or absent ($=\infty$) . When both are present, two sub-cases are formed based on the relative timing of the input and output spikes in the classical STDP manner \cite{bi1998synaptic}. In effect, a synaptic weight is incremented (strengthened) if there is an input spike and it either contributed (Case 1) or can potentially contribute (Case 3) to the output spike; else it is decremented.

The STDP update function either increments the weight by $\Delta$w (up to a maximum of $w_{max}$ = 7), decrements the weight by $\Delta$w (down to a minimum of 0), or leaves the weight un-changed. The $\Delta$ values are defined using Bernoulli random variables (BRVs) with parameterized learning probabilities denoted as B($\mu$) with a descriptive subscript. $F(w)$ is a stabilization function (=$B((w/w_{max})(1-w/w_{max}))$) which makes the weights "sticky" at both ends (0 and 7).

\subsection{Proposed STDP Implementation}
\begin{figure}[t]
        \centering
        \includegraphics[width=3.5in]{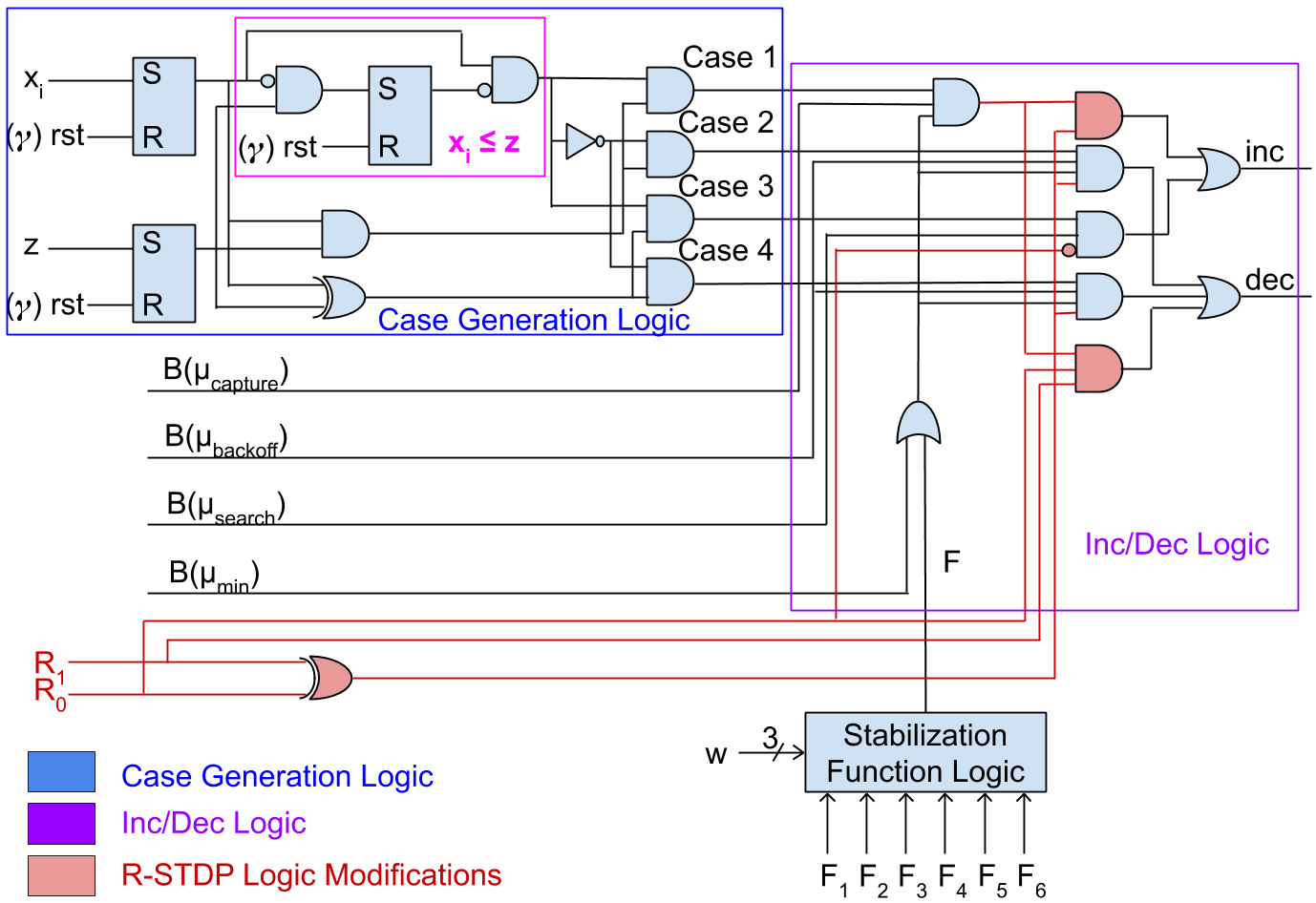}
        \caption{STDP and R-STDP Logic Implementation}
        \label{stdp_logic}
\end{figure}
The proposed STDP logic implementation is shown in Figure \ref{stdp_logic}. It generates 2 control signals (increment/decrement) at the output that feed into the synaptic weight counters described in Figure \ref{pt}. Here, it is assumed that all the required BRVs have been generated from an LFSR network shared by many neurons or columns.
Although detailed LFSR network designs are not considered here, introducing LFSRs into the implementation should be straightforward once their distribution across the multi-layer TNN is determined.
Note that STDP updates (and the associated resets) are performed at the end of a computational cycle (or onset of next gamma clock); inputs for the new computational cycle begin a unit clock cycle later.

The proposed STDP logic implementation can be partitioned into three components.

\subsubsection{Case Generation Logic}
The per-synapse case generation logic compares the synapse's input spiketime (\textit{x}\textsubscript{i}) with its post-synaptic neuron's output spiketime (\textit{z}) and generates 4 control signals corresponding to the 4 cases in Table \ref{stdp}. Case 5 is implicitly invoked when none of the other 4 cases is a 1.
The logic equations implemented for the 4 STDP cases are:

\begin{itemize*}
\item Case 1: $(x_i\leq z).(x_i).(z)$
\item Case 2: $(\overline{x_i\leq z}).(x_i).(z)$
\end{itemize*}

\begin{itemize*}
\item Case 3: $(x_i\leq z).(x_i\oplus z)$
\item Case 4: $(\overline{x_i\leq z}).(x_i\oplus z)$
\end{itemize*}

Note that ($(x_i\leq z)$) is implemented here using a much simpler temporal comparator as opposed to a binary comparator. If \textit{z} arrives prior to \textit{x}, the output is 0; else \textit{x} is allowed to pass. The gate count for this logic is $17p+5$.

\subsubsection{Stabilization Function Logic}
This logic selects 1 BRV from a set of finite BRVs generated by $F(w)$, based on the synaptic weight. For $w_{max}=7$, there are 6 non-zero BRVs to choose from. 
Its logic implementation 
results in $12p$ gates.

\subsubsection{Inc/Dec Logic}
The inc/dec logic assumes 4 BRV inputs from the LFSR network corresponding to the four STDP cases. The \textit{max} operation in Table \ref{stdp} is simply implemented by ‘OR’ing `F' with \textit{min} BRV input. The output of the stabilization logic is used along with the cases from case generation logic to generate \textit{inc} and \textit{dec} outputs.  Its gate count comes out to be just $7p$.

Adding the above gate counts, the STDP logic uses $36p+5$ gates in total. Now, the total gate count for a neuron with \textit{p} synapses, including STDP, can be calculated as below.
\begin{equation}
\label{neuron_w_stdp}
\setlength{\abovedisplayskip}{3pt}
\mathbf{Gates_{neuron\_STDP}=102p+8log_2p+36}
\end{equation}
\subsection{Proposed R-STDP Implementation}
This subsection introduces a variation of our proposed STDP capable of \textit{reinforcement learning} (R-STDP) that uses an external \textit{reward} signal to drive its learning process towards a desired direction.
It involves three forms of reinforcement:
\begin{itemize}[noitemsep, topsep=2pt]
    \item When the column's (non-null) output matches the desired action, \textit{reward} = `1'. It operates as per Table \ref{stdp}; except
    case 3 results in no synaptic weight update.
    \item When the column's (non-null) output does not match the desired action, \textit{reward} = `-1'. 
    Only Case 1 and Case 3 are performed;    
    for Case 1, instead of incrementing the weight, it is decremented.
    \item When the column produces no output, i.e., no neuron spikes, \textit{reward} = `0' and only Case 3 operates.
\end{itemize}


In effect, desired behavior is reinforced and undesirable behavior is repressed. 
Note that R-STDP is still applied locally to each neuron. However, the context for R-STDP is at the column or layer level due to the global \textit{reward} signal.

The logic modifications required for R-STDP are rather minimal and straightforward as highlighted in Figure \ref{stdp_logic}. \textit{reward} is a 2-bit signal (which encodes `-1', `0' and `1' as `11', `00' and `01' respectively). Unsupervised STDP is invoked when \textit{reward} is `10'.
The total gate count for a neuron that implements R-STDP is given below:
\begin{equation}
\label{neuron_w_rstdp}
\setlength{\abovedisplayskip}{3pt}
\mathbf{Gates_{neuron\_RSTDP}=106p+8log_2p+36}
\end{equation}
%
To the best of our knowledge such gate-level and hardware-efficient implementations of STDP and R-STDP have not been presented or published before.

\section{Column Design and Implementation}
\label{column_sec}
In this section, the column microarchitecture is defined and a scalable implementation is presented. As shown in Figure \ref{fig4a_new}, a \textit{column} is a stack of excitatory neurons operating in parallel, followed by a winner-take-all (WTA) inhibition across the neurons. STDP is performed after WTA. Columns are used as functional building blocks for TNNs as shown in Figure \ref{fig2_new}.

\subsection{Column Operation}
In this subsection, we briefly explain the functionality of a typical column before moving on to its implementation in the following subsections. Figure \ref{fig4b_new} illustrates the operation of an example column with 8 RNL neurons and a total of 64 synaptic weights (each neuron has 8 input synapses and a threshold value of 8). Maximum synaptic weight value is 7. Every cross point in the synaptic crossbar is a synapse with its weight value shown inside the circle. Absence of values indicates a weight of 0. Neuron 1 (driving output $z_1$), has only one of its inputs ($x_5$) driving a synapse with weight = $w_{max}$, which is lower than the threshold. Hence, it doesn't produce an output spike (t = $\infty$). This value ($\infty$) is shown as the input to 1-WTA inhibition in Figure \ref{fig4b_new}. Neuron 4 ($z_4$) receives input spikes on three inputs with weights of $w_{max}$; hence its body potential will cross the threshold at t = 2. The output $z_4$ is the first output spike, so it is selected as the ``winner'' by WTA inhibition. Hence, $z_4$ = 2; all the other $z_i$ = $\infty$.
\begin{figure}[t]
        \centering
        \includegraphics[width=3.2in]{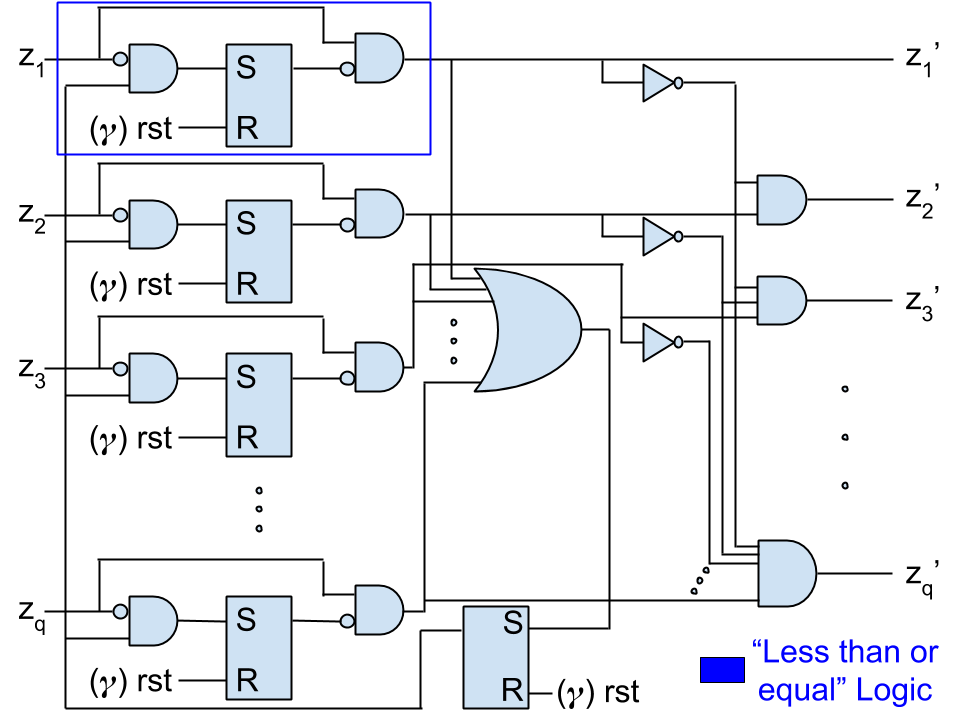}
        \caption{WTA Inhibition for a Column of \textit{q} Neurons}
        \label{li}
\end{figure}
\subsection{Winner-Take-All (WTA) Inhibition}
Winner-take-all (WTA) lateral inhibition selects the first (or first
\textit{k} as in \textit{k}-WTA) spiking neuron and allows its output spike to pass through intact, while nullifying the outputs of the other neighboring neurons. Figure \ref{li} shows the logic diagram for 1-WTA inhibition across the \textit{q} neurons in a column.
The inhibition operation is performed by a latch-based less-than-or-equal temporal comparison unit (same as in the case generation logic).
The first spike is found through a big ‘OR’ gate, or a tree of small OR gates, (performing a temporal 'min' function) and is fed back through a latch which holds the signal at 1 until the next gamma cycle. Any input pulse coming to the latch after this signal is blocked, so only the first spikes are passed.
Tie breaking is implemented as a priority-based logic that selects the first spiking neuron with the lowest index. Gate count for 1-WTA (upper bound) is $8q+q^2$.
\subsection{Proposed Column Implementation}
As shown in Figure \ref{fig4a_new}, each column contains \textit{q} excitatory neurons and a synaptic crossbar connecting the \textit{p} inputs to the \textit{q} neurons via \textit{p}x\textit{q} synapses. A column supports unsupervised learning via STDP or supervised learning via R-STDP at each of those synapses. A column also supports WTA lateral inhibition to assist in convergence of synaptic weights.

A single (\textit{p}x\textit{q}) column with \textit{p} synaptic inputs and \textit{q} excitatory neurons, supported by STDP or R-STDP and WTA becomes a fully operational TNN, capable of performing inferencing and online continuous learning. Columns can also be used as building blocks for creating larger TNNs by stacking multiple columns to form a multi-column layer, as well as by cascading multiple layers into a large multi-layer TNNs as illustrated in Figure \ref{fig2_new}.

Equations (\ref{column_w_stdp}) and (\ref{column_w_rstdp}) below provide the total gate counts for a \textit{p}x\textit{q} column 
with STDP and R-STDP, respectively. These two equations effectively characterize the hardware implementation complexity (in total gate count) of an arbitrary column with \textit{q} neurons, each with \textit{p} synapses, supporting unsupervised and supervised learning paradigms.
\begin{equation}
\label{column_w_stdp}
\setlength{\abovedisplayskip}{3pt}
\mathbf{Gates_{col\_STDP}=102pq+8qlog_2p+44q+q^2}
\end{equation}
\vspace{-18pt}
\begin{equation}
\label{column_w_rstdp}
\setlength{\abovedisplayskip}{3pt}
\mathbf{Gates_{col\_RSTDP}=106pq+8qlog_2p+44q+q^2}
\end{equation}


\section{Column Implementation Evaluation}
Scalable designs of neurons and columns have been implemented in Verilog; synthesis results are generated based on a 45nm standard cell library using Synopsys tools. This section presents evaluation of these designs.

\subsection{Methodology}
\label{method_sec}
Two types of evaluation, namely gate-level and circuit-level, are performed to determine area (A), (neuron) critical path delay (D), (column) computation time (T) and power (P) for the proposed neuron and column designs.

In gate level evaluation, we derive equations for A, D, T and P based on gate count and number of signal transitions, parameterized in terms of number of neurons (\textit{q}) and number of synapses per neuron (\textit{p}). The procedure is as follows: 1) Gate count is used as a surrogate for A. 2) Total number of gates in the critical path is used for D. T is calculated using the formula $T = (t_{max}+w_{max}+1) * D$. For a b-bit time window, the first and last spikes can be separated by up to $t_{max}$ clock cycles ($t_{max}=2^b-1$). After the last input spike arrives, it can take up to $w_{max}-1$ more cycles for the RNL response function to reach its peak, 1 cycle for restoring the weights and a final cycle for STDP update, hence $w_{max}+1$ cycles are added. 3) Gate count is used for estimating static power, and the number of gate transitions for dynamic power consumption.
`AND' gate can be taken as the reference gate for the following gate-level evaluations.
%

Circuit level results are obtained from post-synthesis output generated by Synopsys Design Compiler using open source 45 nm Nangate standard cell library \cite{knudsen2008nangate}. We use the low power process corner for synthesis. Area, power and critical path delay are obtained directly from the Synopsys tool, and computation time is derived by multiplying the critical path delay by 15. The post-synthesis designs have been verified functionally both in Python and Verilog using Synopsys VCS.
\begin{scriptsize}
\begin{table}[t]
  \centering
  \scalebox{1.10}{
  \begin{tabular}{|c|r|c|c|c|}
    \hline
    \textbf{Syn-} & \textbf{Gate} & \textbf{Area} & \textbf{Critical Path} & \textbf{Power}\\
    \textbf{apses} & \textbf{Count} & [mm\textsuperscript{2}] & \textbf{Delay} [ns] & [mW]\\
    \hline
    \hline
    64 & 6,471 & 0.0065 & 1.93 & 0.031 \\
    \hline
    128 & 12,859 & 0.0129 & 2.16 & 0.062 \\
    \hline
    256 & 25,673 & 0.0258 & 2.41 & 0.124 \\
    \hline
    512 & 51,258 & 0.0515 & 2.64 & 0.249 \\
    \hline
    1024 & 102,432 & 0.1030 & 2.82 & 0.497 \\
    \hline
  \end{tabular}
  }
  \caption{Area, Delay and Power (ADP) Values: (in 45 nm CMOS) for a neuron with STDP relative to synapse counts.}
  \label{nresults}
\end{table}
\end{scriptsize}
\renewcommand{\tabcolsep}{2pt}
\begin{scriptsize}
\begin{table}[t]
  \centering
  \resizebox{9cm}{1.111cm}{
  \begin{tabular}{|c|c|c|}
    \hline
    \textbf{Metrics} & \textbf{Neuron} & \textbf{Column}\\
    \hline
    \hline
    A & $102p+8log_2p+36$ & $102pq+8qlog_2p+44q+q^2$ \\
    \hline
    D / T & $6log_2p+4$ & $90log_2p+60$ \\
    \hline
    P\textsubscript{static} & $102p+8log_2p+36$ & $102pq+8qlog_2p+44q+q^2$ \\
    \hline
    P\textsubscript{dynamic} & $204p+185log_2p+241$ & $204pq+185qlog_2p+257q+2q^2$ \\
    \hline
  \end{tabular}
  }
  \caption{Gate Level Equations: for A, D, T and P for a neuron with \textit{p} synapses and a \textit{p}x\textit{q} column implementing STDP. D is for neuron and T is for column.}
  \label{gateeval}
\end{table}
\end{scriptsize}
\vspace{-13pt}
\subsection{Neuron Gate Level Evaluation}
The equations in Table \ref{gateeval} can be used as formulae to scale A, D and P for arbitrary number of synapses for a neuron. Area and static power are straightforward. Critical path in a neuron is the path from the FSM to the output of the accumulator (shown by red arrow in Figure \ref{acc}); the number of gates in this path is used to estimate D. All gates in the FSM undergo at most 2 signal transitions during the 8 states, since the RNL pulses have to be consecutive in time, thereby resulting in only 2 transitions, one at the beginning and one at the end. Every gate except the gates in the last stage in the accumulator (marked by purple box in Figure \ref{acc}) undergoes at most 2 transitions from incoming pulses from FSMs. For an estimate of worst-case power consumption, the last stage gates are assumed to undergo 15 transitions in 15 clock cycles.

Gate level equations derived above indicate near-linear scaling of area and power, and logarithmic scaling of delay with respect to the number of synapses. We verify this with the post-synthesis results for 45 nm CMOS generated by Synopsys Design Compiler in the next subsection.

\subsection{Neuron Circuit Level Evaluation}
\subsubsection{Gate Complexity Breakdown}
\begin{figure}[t]
    \centering
    \begin{subfigure}[b]{0.25\textwidth}
        \centering
        \includegraphics[width=\textwidth,height=3.85cm]{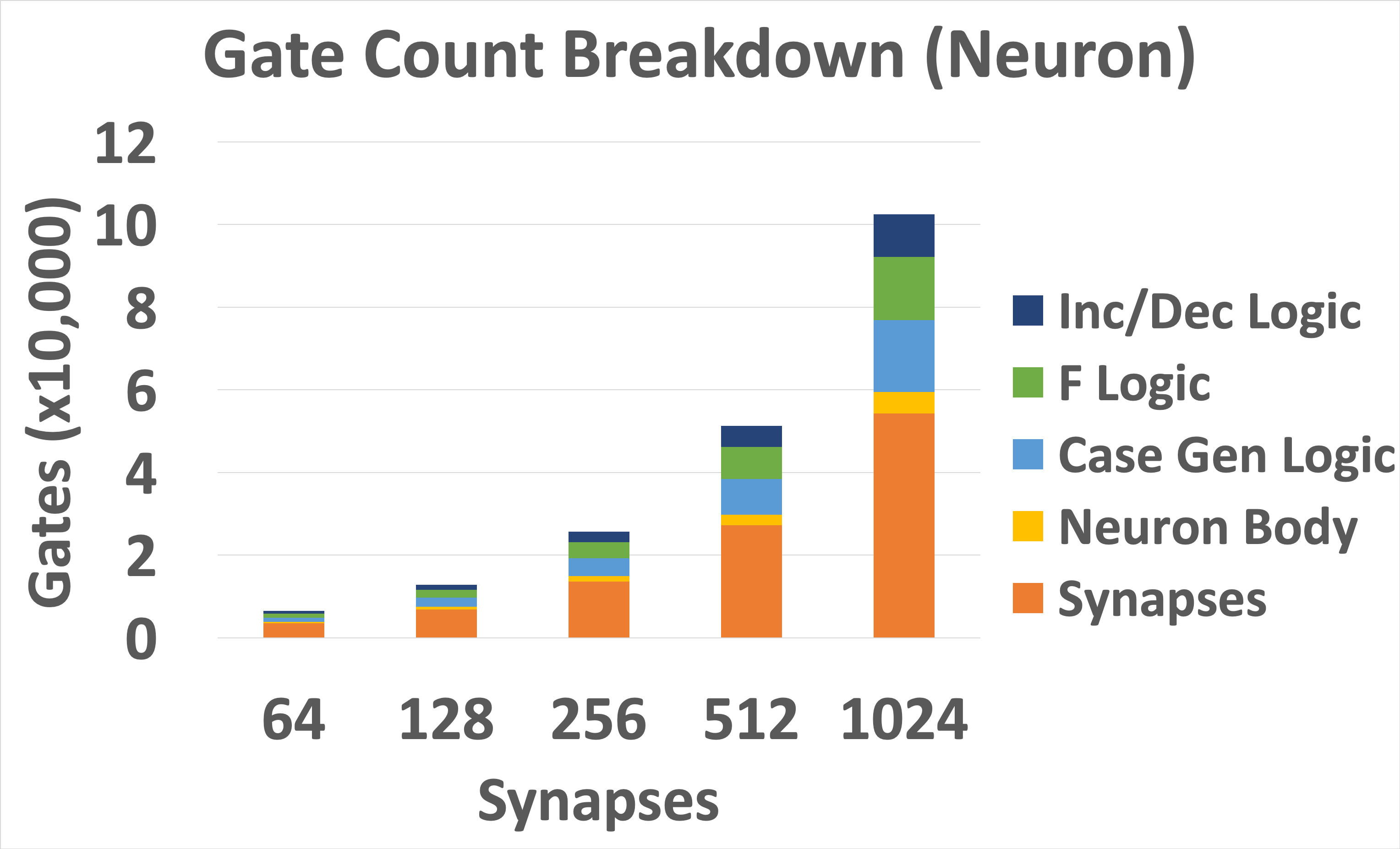}
        \caption{Neuron Gate Count}
        \label{gate_neuron}
    \end{subfigure}%
    %
    \begin{subfigure}[b]{0.25\textwidth}
        \centering
        \includegraphics[width=\textwidth,height=3.85cm]{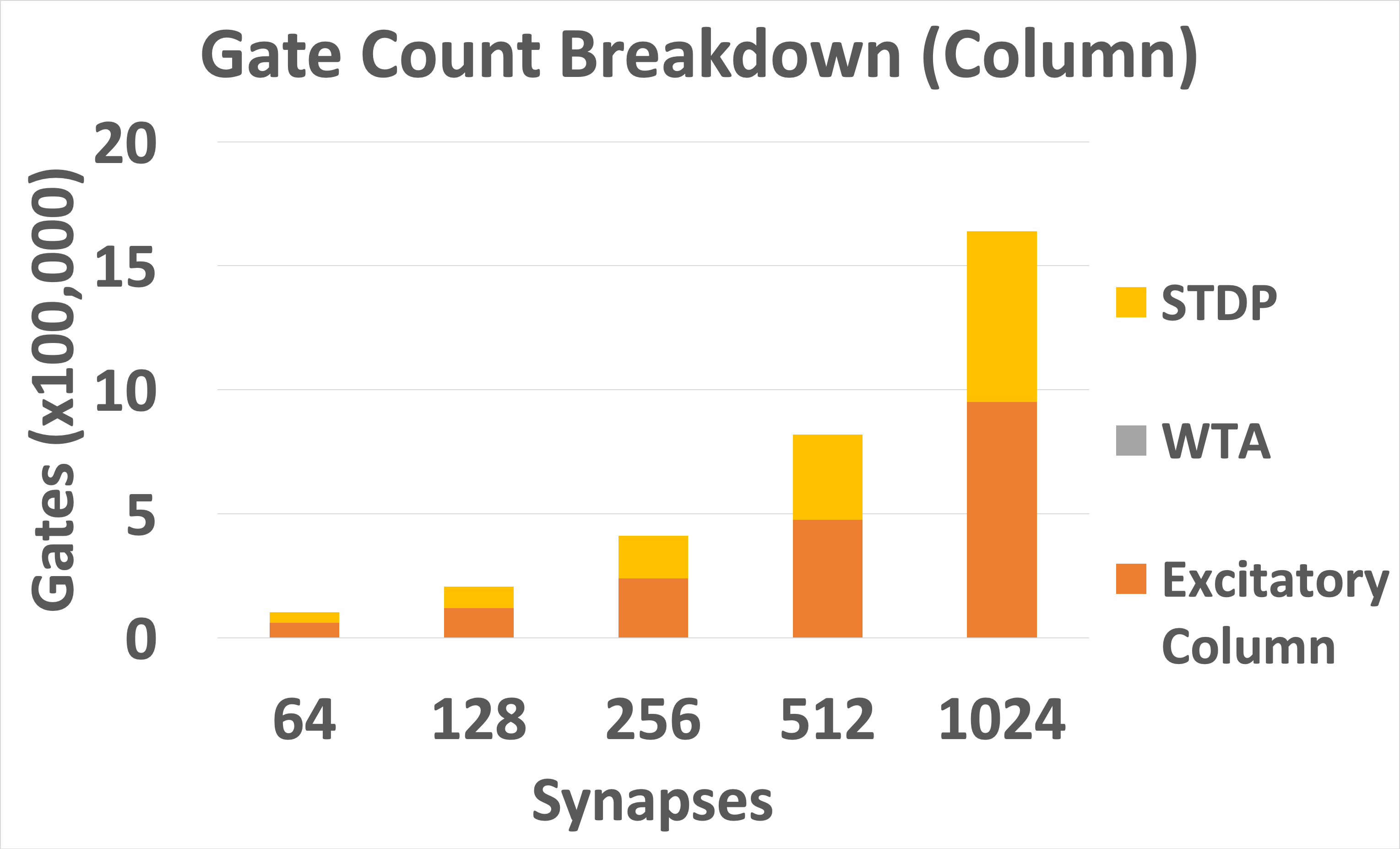}
        \caption{Column Gate Count}
        \label{gate_column}
    \end{subfigure}
    \caption{Gate Count Breakdown: for a neuron and a column of 16 neurons, varying synapses from 64 to 1024.}
\end{figure}
Figure \ref{gate_neuron} shows the breakdown for various components of the neuron in terms of gate count, scaled across synapse count from 64 to 1024. We observe a near-linear scaling of total gate counts relative to the number of synapses. The gate count for the neuron body is quite small whereas the synapses incur the highest gate counts. Synapses constitute almost 50\% of the entire neuron complexity and STDP logic 40\% while the neuron body accounts for the remaining 10\%. A neuron with 1024 synapses has a total of 102,432 gates.

\subsubsection{ADP Complexity (45 nm CMOS)}
Table \ref{nresults} shows the A, D and P values for a neuron with STDP-programmable synapses, for 45 nm digital CMOS technology. It also includes gate count generated from the post-synthesis output of Design Compiler. A temporal neuron with 1024 synapses consumes about 0.1 mm\textsuperscript{2} area and 0.5 mW power. Its critical path delay of 2.82 ns maps to a maximum clock frequency of 354.6 MHz. This is the maximum rate at which the actual hardware clock \textit{aclk} can be clocked. The near-linear scaling of A and P, and logarithmic scaling of D, relative to synapse counts, can be verified based on the values in the table. The gate count from post-synthesis results in Table \ref{nresults} can be easily corroborated by inserting appropriate values for \textit{p} in Equation (\ref{neuron_w_stdp}).
%

\subsection{Column Gate Level Evaluation}
A, T and P gate level equations for a \textit{p}x\textit{q} column having \textit{q} neurons, each with \textit{p} synapses, implementing unsupervised STDP are given in Table \ref{gateeval}. These equations (area/power) can be easily adapted for R-STDP using Equation (\ref{column_w_rstdp}).

Time for a single computational cycle (T) or the gamma cycle can be derived as mentioned in Section \ref{method_sec} using $w_{max}=t_{max}=7$. This is because a column's critical path is the same as that for a neuron. For estimating dynamic power consumption, each gate in WTA inhibition logic is assumed to have 2 transitions (including 1 after reset).
%



\subsection{Column Circuit Level Evaluation}

\renewcommand{\tabcolsep}{6pt}
\begin{scriptsize}
\begin{table}[t]
  \centering
  \scalebox{1.0}{
  \begin{tabular}{|c|c|r|c|c|c|}
    \hline
    & \textbf{Synapses x} & \textbf{Gate} & \textbf{Area} & \textbf{Comp.} & \textbf{Power}\\
    & \textbf{Neurons} & \textbf{Count} & [mm\textsuperscript{2}] & \textbf{Time} [ns] & [mW]\\
    \hline
    \hline
    \multirow{3}{*}{STDP} & 64 x 8 & 51,824 & 0.05 & 28.95 & 0.25 \\
    \cline{2-6}
    & 128 x 10 & 128,658 & 0.13 & 32.40 & 0.62 \\
    \cline{2-6}
    & 1024 x 16 & 1,639,020 & 1.65 & 42.30 & 7.96 \\
    \hline
    \multirow{3}{*}{R-STDP} & 64 x 8 & 54,384 & 0.05 & 28.95 & 0.26 \\
    \cline{2-6}
    & 128 x 10 & 135,058 & 0.14 & 32.40 & 0.65 \\
    \cline{2-6}
    & 1024 x 16 & 1,720,940 & 1.75 & 42.30 & 8.36 \\
    \hline
  \end{tabular}
  }
  \caption{Area, Computation Time, and Power Consumption Values: (in 45 nm CMOS) for three column sizes of: 64x8, 128x10, 1024x16, implementing STDP and R-STDP learning rules. R-STDP incurs minimal overhead over STDP.}
  \label{cresults}
\end{table}
\end{scriptsize}
Figure \ref{gate_column} shows the gate count breakdown for the three components of a TNN column with 16 neurons, varying synapse count for each neuron from 64 to 1024. Excitatory column constitutes just above 60\% of the entire column complexity while STDP logic incurs close to 39\%. Interestingly, WTA lateral inhibition incurs  negligible gate count. As can be seen from Figure \ref{gate_column}, total column gate count again exhibits a near-linear scaling relative to the total synapse count. In Table \ref{cresults}, we present 45 nm CMOS results for three column configurations with both STDP and R-STDP learning rules: 1) a small 64x8 column, 2) a medium 128x10 column, and 3) a large 1024x16 column. The \textit{gamma} cycles for the smallest column and the largest column are 28.95 ns (34.54 MHz) and 42.3 ns (23.64 MHz) respectively. Smallest column with STDP consumes about 0.05 mm\textsuperscript{2} area and 0.25 mW power, whereas the largest column has an area and power footprint close to 1.7 mm\textsuperscript{2} and 8 mW respectively. Note that R-STDP increases area and power by only 5\% relative to STDP.

\section{A Multi-Layer TNN Prototype}

This section presents a TNN prototype design based on our TNN microarchitecture in Figure \ref{fig2_new}, using the scalable multi-neuron columns and multi-column layers as building blocks. This 2-layer TNN prototype has one unsupervised layer followed by a supervised layer as shown in Figure \ref{tnn_prototype}, with dimensions: \textbf{TNN\{[625x(32x12)]+[625x(12x10)]\}}.

\begin{figure}[t]
    \centering
    \includegraphics[width=3.5in]{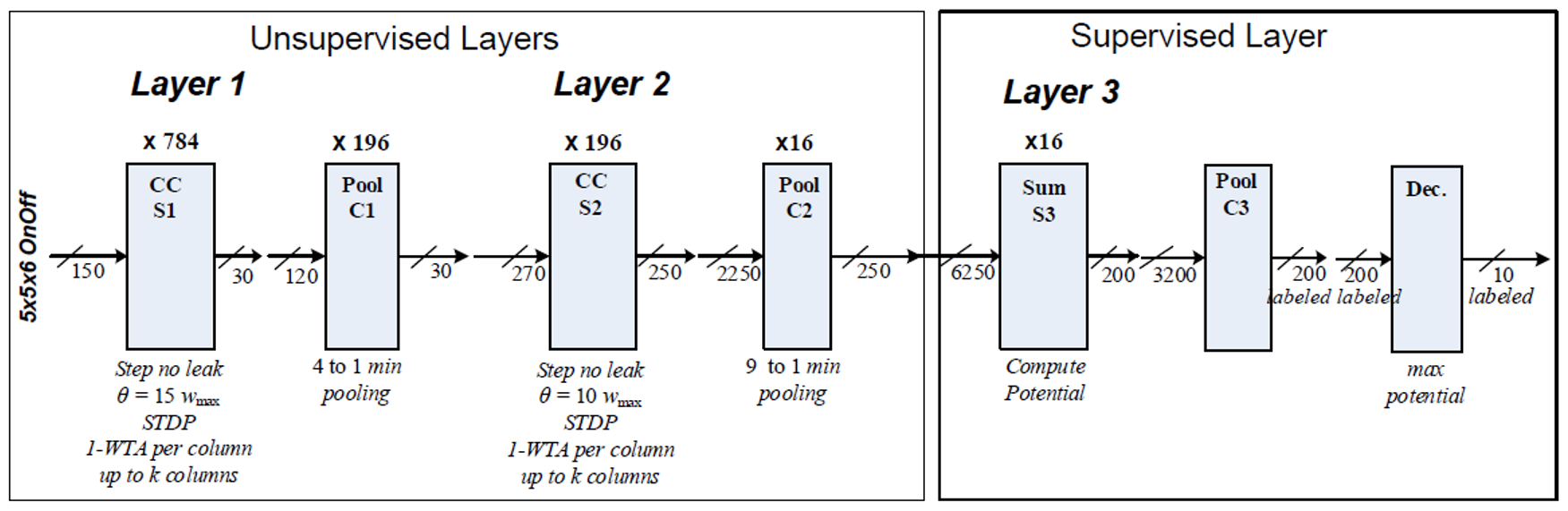}
    \caption{Baseline Model: consisting of two unsupervised layers and one supervised layer, from Mozafari et al. \cite{mozafari2019bio}.}
    \label{thorpe_model}
\end{figure}

\begin{figure}[t]
    \centering
    \includegraphics[width=2.7in]{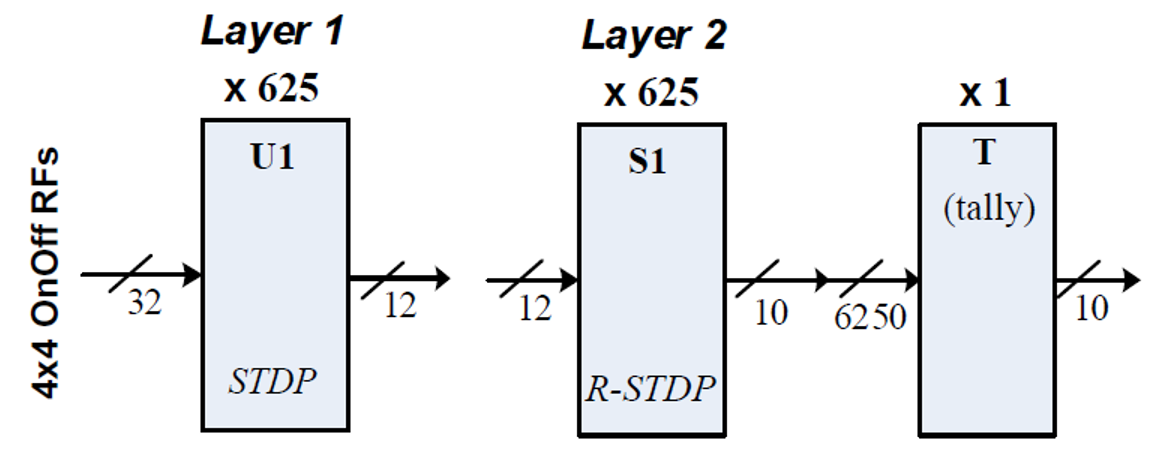}
    \caption{Proposed TNN Prototype: consisting of one unsupervised layer and one supervised layer with dimensions: \textbf{TNN\{[625x(32x12)]+[625x(12x10)]\}}.}
    \label{tnn_prototype}
\end{figure}

We compare this prototype network against the 3-layer baseline network shown in Figure \ref{thorpe_model} from Mozafari et al. \cite{mozafari2019bio}. This baseline network represents the current state-of-the-art TNN model. As was used in\cite{mozafari2019bio} for the baseline network, we also use the MNIST data set \cite{lecun1998gradient} for our benchmarking. MNIST data set of 28x28 images is widely used for image classification.
We compare the two networks based on MNIST training time, accuracy, and total synapse count. Note that the goal here is not to propose a state-of-the-art TNN accelerator for MNIST, but merely to demonstrate the potential of TNNs for future TNN-based sensory processing units.

We also present hardware metrics for our TNN prototype, including total gate count, die area, compute time, and power consumption, with technology scaling from 45 nm to 7 nm.


\subsection{Evaluation of Our TNN Prototype}

\textit{Baseline Network Model:} See Mozafari et al. \cite{mozafari2019bio}. The baseline network as shown in Figure \ref{thorpe_model}, consists of two unsupervised layers followed by a supervised layer for classification. For direct comparison, we convert this network from the original feature-map organization used by the authors to an equivalent TNN column organization. 

Given a convolution layer in the baseline model, the number of equivalent TNN columns can be determined using the input feature map size, kernel window size and the stride. The number of neurons per column (\#outputs) is equal to the number of output feature maps, since each neuron in a column learns a distinctive feature. The number of synapses per neuron in each column (\#inputs) is given by the number of input feature maps and the kernel window size. Essentially, the column organization is a  structural transpose of the feature map organization with the same total synapse count.

%
%
Parameters for all the layers of the baseline network are shown in Figure \ref{thorpe_model}, after conversion to the equivalent column organization.
Each layer has a pooling sub-layer attached. This network uses 5x5 receptive fields (RF) and DoG filtering with 6 channels.
The complexity of each layer is based on the synapses count (\#synapses), which can be calculated using dimension (\#inputs x \#outputs) of each column, and the number of columns in each layer (\#columns); see Table \ref{tnneval}. 

The entire 3-layer baseline network employs over 36 million synapses. Although not explicitly constructed as an online system, it can be adapted to operate online in a straightforward way. The first layer is trained with 100K input patterns, the second layer with 200K, and the supervised layer requires 40M. The best MNIST accuracy achieved after 40M training inputs is 97\%. We have validated these results in the SpykeTorch \cite{mozafari2019spyketorch} based functional simulator.
\begin{scriptsize}
\begin{table}[t]
  \centering
  \scalebox{0.98}{
  \begin{tabular}{|c|c|c|c|c|r|}
    \hline
     &  & \textbf{\#} & \textbf{\#} & \textbf{\#} & \textbf{\#} (1000's)\\
     &  & \textbf{Inputs} & \textbf{Outputs} & \textbf{Columns} & \textbf{Synapses}\\
    \hline
    \hline
    & Layer 1 & 150 & 30 & 784 & 3,528 \\
    \cline{2-6}
    Baseline & Layer 2 & 270 & 250 & 196 & 13,230 \\
    \cline{2-6}
    Model & Layer 3 & 6250 & 200 & 16 & 20,000 \\
    \cline{2-6}
    &  &  &  & \cellcolor{yellow} Total & \cellcolor{yellow} 36,758 \\
    \hline
    \multirow{3}{10mm}{\hspace{1mm} TNN Prototype} & U1:Layer 1 & 32 & 12 & 625 & 240 \\
    \cline{2-6}
    & S1:Layer 2 & 12 & 10 & 625 & 75 \\
    \cline{2-6}
    &  &  &  & \cellcolor{yellow} Total & \cellcolor{yellow} 315 \\
    \hline
  \end{tabular}
  }
  \caption{Complexity Comparison: 3-layer baseline model vs. our 2-layer TNN prototype in terms of total synapse counts.}
  \label{tnneval}
\end{table}
\end{scriptsize}
%

\textit{Prototype TNN Network:} This network, shown in Figure \ref{tnn_prototype}, consists of two layers, one unsupervised and one supervised. It uses 4x4 receptive fields (RF) with On/Off encoding and stride of 1 across the 28x28 image resulting in 25x25 RFs. Hence, Layer 1 has 625 columns (one for each RF) of size (32x12). Layer 2 has the same number of columns but each of size (12x10), and is essentially a supervised voting layer that generates 625 votes (1 or 0) for each label. The Tally (T) sub-layer of Layer 2, consists of 10 adder trees, each with 625 inputs and determines the label with the maximum votes. 

We implement a fully functional model of our TNN prototype network using PyTorch \cite{paszke2019pytorch} and therefore can make direct comparison with the baseline network at the functional level. Since \cite{mozafari2019bio} does not provide a direct hardware implementation, we use synapse count (dominant factor in hardware complexity) for comparison. From Table \ref{tnneval}, the total synapse count for the prototype network is only 315,000. This is two orders of magnitude less than the baseline network (36M synapses), demonstrating the  efficiency of our proposed TNN microarchitecture.
The MNIST accuracy achieved by the simple two-layer network is 93\%.
(A deeper network, beyond the scope of this paper, can achieve 98\%.) Our 93\% accuracy is achieved with less than 30K training inputs (vs. 40M with the baseline), exhibiting very quick weight convergence.

\begin{scriptsize}
\begin{table}[t]
  \centering
  \scalebox{1.10}{
  \begin{tabular}{|c|c|r|r|r|}
    \hline
    \textbf{Tech.} & \textbf{Transistor} & \textbf{Area} & \textbf{Comp.} & \textbf{Power}\\
    \textbf{Node} & \textbf{Density} & [mm\textsuperscript{2}] & \textbf{Time} [ns] & [mW]\\
    \hline
    \hline
    45 nm & 4 MT/mm\textsuperscript{2} & 32.61 & 43.05 & 154.36 \\
    \hline
    28 nm & 10 MT/mm\textsuperscript{2} & 13.04 & 27.23 & 61.74 \\
    \hline
    16 nm & 22 MT/mm\textsuperscript{2} & 5.93 & 18.36 & 28.06 \\
    \hline
    10 nm & 46 MT/mm\textsuperscript{2} & 2.84 & 12.70 & 13.42 \\
    \hline
    \rowcolor{yellow} 7 nm & 85 MT/mm\textsuperscript{2} & 1.54 & 9.34 & 7.26 \\
    \hline
  \end{tabular}
  }
  \caption{Technology Scaling (from 45nm to 7nm) of the TNN Prototype (complexity: 32M gates or 128M transistors).}
  \label{tnn_adp}
\end{table}
\end{scriptsize}

\begin{figure}[t]
    \centering
    \begin{subfigure}[b]{0.23\textwidth}
        \centering
        \includegraphics[width=\textwidth]{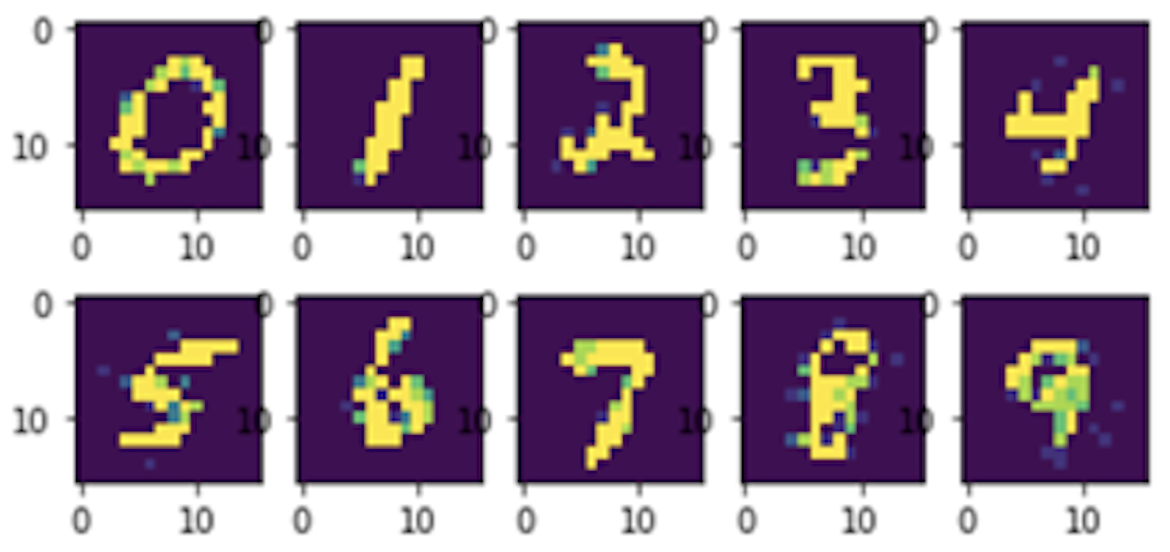}
        \caption{Trained for Labels 0 - 9}
        \label{fig_14a}
    \end{subfigure}%
    \begin{subfigure}[b]{0.23\textwidth}
        \centering
        \includegraphics[width=\textwidth]{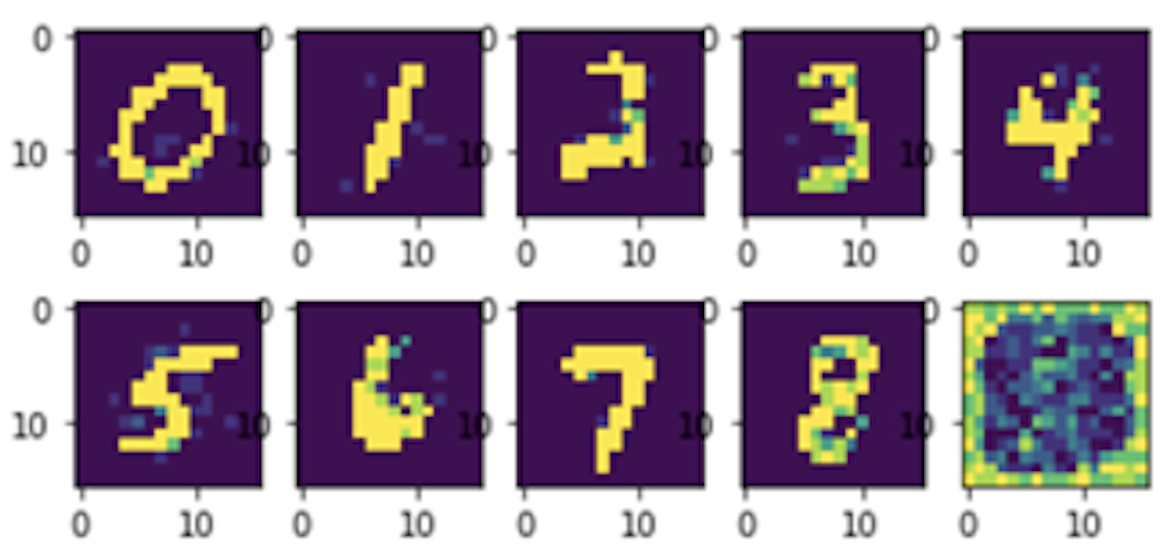}
        \caption{Trained for Labels 0 - 8}
        \label{fig_14b}
    \end{subfigure}
    \label{w_num}
    \caption{Synaptic weight matrices converge to image centers resembling MNIST labels in just 10,000 samples.}
\end{figure}
\begin{figure}
        \centering
        \includegraphics[width=3.55in]{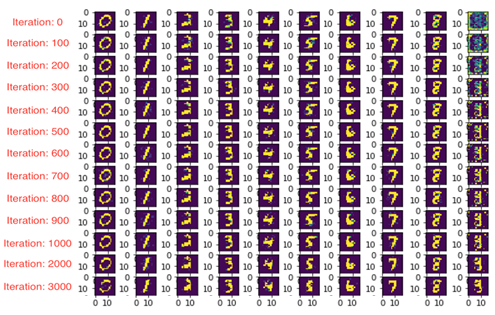}
        \caption{Online Incremental Learning: STDP learns a previously unseen input number '9' within 500 testing samples.}
        \label{inc_learn}
\end{figure}

\subsection{Online Image Classification}

In contrast to the typical epoch-based, back propagation training methods, STDP is an online training method that consumes and processes the inputs in a streaming manner and amenable to online real-time applications.
Our goal is not to compete with state-of-the-art performance on MNIST, but just using it as a benchmark to validate the functionality and demonstrate the efficiency of TNNs. Nevertheless, based on our experiments with this benchmark for a single column, several interesting capabilities of TNNs can be observed.
\begin{enumerate}[noitemsep, topsep=2pt]
    \item \textit{Online Classification via Centroid Formation}: Figure \ref{fig_14a} shows the synaptic weights converged to the 10 class centroids, which resemble the corresponding digits. 
    \item \textit{Fast Training Convergence}: The synaptic weights in Figure \ref{fig_14a} and Figure \ref{fig_14b} converged after approximately 10,000 training samples, which implies that TNNs can learn very quickly and can generalize from  small datasets.
    \item \textit{Online Incremental Learning}: In this experiment, R-STDP training is first performed with only 9 classes (0 to 8) by hiding the label '9', resulting in the converged weights shown in Figure \ref{fig_14b}. Then the label '9' is introduced in the input sequence to illustrate the ability to dynamically learn  a previously unseen class. As shown in Figure \ref{inc_learn}, the rightmost synaptic weight matrix converges to label '9' after only about 500 testing samples.
\end{enumerate}
TNNs have the potential for both incremental and continuous learning. Online incremental learning enables a TNN to adapt to new input data not seen before during the original (offline) training. Continuous learning allows a TNN to keep learning and improving its accuracy concurrently with inference.

\subsection{Hardware Complexity and Technology Scaling}

Based on our results from earlier sections, we can assess the hardware complexity, performance, and power of the prototype network
(or any other arbitrary multi-layer TNN network). 
Using the gate-level implementation of our scalable column design and Equation \ref{column_w_stdp} and Equation \ref{column_w_rstdp}, we can compute the gate-level complexity for the prototype TNN as: 32M gates or 128M transistors. Further breakdown of the gate counts for the layers are: 24.12M (U1), 7.91M (S1), and 31.25K (T).

Using 45 nm standard cell library and transistor density of 4 MT/mm\textsuperscript{2}, we have 32.61 mm\textsuperscript{2} die area, 43.05 ns compute time, and 154.36 mW power consumption for the prototype TNN as shown in the first row of Table \ref{tnn_adp}.
Using a common approach based on the scaling of transistor density from one process node to the next, we can estimate the die area, compute time, and power consumption for our TNN prototype as we scale the CMOS process from 45 nm down to 7 nm.

Technology scaling for area and power can be performed by multiplying them by the ratio of transistor densities between the source and target nodes. For critical path delay, we use the square root of the above ratio. 
As can be seen in Table \ref{tnn_adp}, across the five technology nodes, ranging from 45nm to 7nm, area and power for each subsequent generation are scaled by a factor of 0.4, 0.45, 0.48 and 0.54, respectively. The critical path delay scales with the square root of these scaling factors, namely, 0.63, 0.67, 0.69 and 0.73, respectively.

With the current 7 nm CMOS process, the entire 2-layer TNN prototype takes up only 1.54 mm\textsuperscript{2} of die area, consumes just 7.26 mW, and incurs only 9.34ns for each compute cycle. Compared to 7 nm SoCs in high-end smartphones, with typical die sizes of over 100 mm\textsuperscript{2}, this is less than 2\% of the mobile SoC die area, and less than 1\% of its 1W power budget. Note that our TNNs do not perform any matrix multiplication. There are no MAC units or SRAM arrays. This is a whole new way of implementing online sensory processing units. 

\section{Concluding Remarks and Future Research}

\begin{figure}[t]
    \centering
    \includegraphics[width=3.3in]{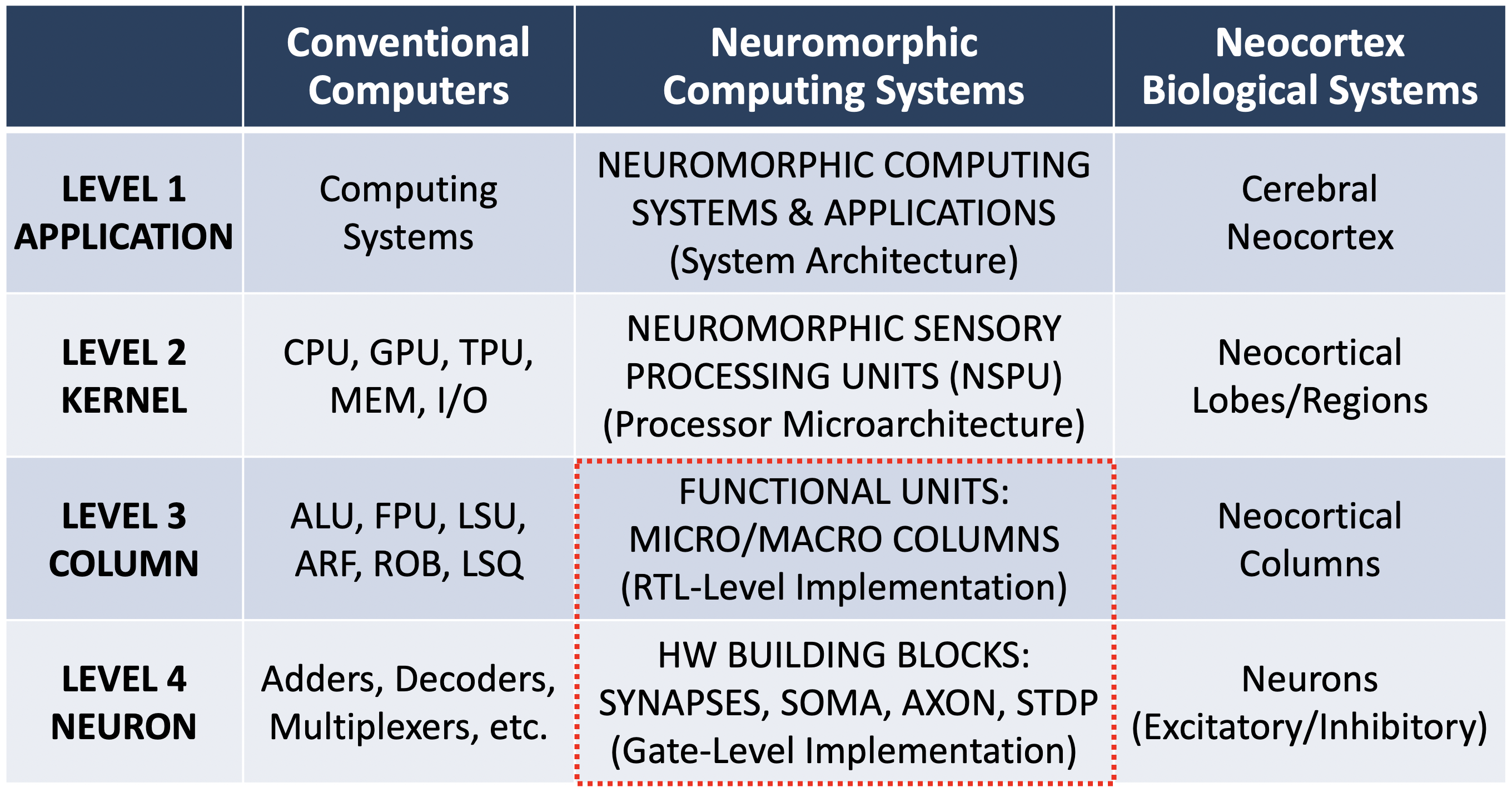}
    \caption{Neuromorphic Computing Abstraction Levels.}
    \label{neuro_abst}
\end{figure}

We take a bottom up approach by examining the hierarchical organization of biological neocortex systems to formulate an analogous hierarchical organization for TNNs. Based on this approach, we propose the neuromorphic abstraction levels shown in Figure \ref{neuro_abst}. This paper focuses on Levels 3 and 4.

Two key questions: "Can such TNNs perform useful functions?" and "Can such TNNs be implemented using standard CMOS?" The work by Mozafari et al. \cite{mozafari2019bio} contributes towards the first question. In this paper, we focus on and address the second question. We implement a scalable column, with STDP and R-STDP, in Verilog, and obtain the post-synthesis results for the 45 nm CMOS process. We can use these two column types to create two layer types for either unsupervised (STDP) or supervised (R-STDP) learning, as shown in Figure \ref{fig5_new}.

The key contributions of this paper include the TNN microarchitecture and implementation model embodied in Equations \ref{column_w_stdp} and \ref{column_w_rstdp}, and Tables \ref{nresults}, \ref{gateeval}, and \ref{cresults}. Equations \ref{column_w_stdp} and \ref{column_w_rstdp} characterize the gate level complexity of TNN columns, from which we can compute the complexity of any multi-column layer (with STDP or R-STDP) and any multi-layer TNN. Table \ref{gateeval} provides process-independent characteristic equations (based on our gate level designs) for die area (A), delay/compute time (D/T), and power consumption (P). Tables \ref{nresults} and \ref{cresults} provide actual data on A, D/T, and P for the neuron and column models for the 45 nm process node. Based on these two tables, we can estimate these metrics for any multi-layer TNN in 45 nm CMOS. Using standard process scaling parameters, we can also estimate these metrics for any TNN design at any process node, as illustrated in Table \ref{tnn_adp}. 

Some key observations from this work include:
\begin{enumerate}[noitemsep, topsep=2pt]
    \item About 50\% of a neuron's complexity comes from the synaptic weights, 40\% from STDP and the remaining 10\% the neuron body.
    \item A column's complexity is mainly determined by its excitatory column's size and configuration, and scales near-linearly with respect to the product of the neuron count and the synapse count.
    \item Area and power for a neuron and a column scale linearly with the total number of synapses \textit{p}x\textit{q}, whereas critical path delay scales logarithmically with the number of synapses per neuron \textit{p}. Since the neuron body mainly determines the critical path delay, the number of neurons \textit{q} in a column has minimal affect on the critical path in a column or even in a multi-layer TNN. 
\end{enumerate}

This work represents only an initial step in a very promising path for follow up research. We take a bottom up approach, starting with Levels 3 and 4 of the hierarchy of neuromorphic abstraction levels in Figure \ref{neuro_abst}. We plan to move up into Levels 1 and 2 in our future research.

We believe that our TNN microarchitecture and direct CMOS implementations can facilitate the design of highly energy efficient, online, and edge-native, sensory processing units capable of incremental and continuous learning that can be incorporated into mobile and edge computing systems as well as always-on IoT sensor processing devices.
The CMOS implementation results in this paper should be viewed as a first opportunistic attempt, using readily-available and existing design methods and tools. There are plenty of potential improvements and new innovations, in both TNN designs and TNN design frameworks, 
waiting to be harvested.

Current computing demand for AI/ML workloads is increasing at the rate of 10x per year or doubling every 3.4 months \cite{openai}. Moore's law, at best, is only doubling every 2 years. The gap between the increasing computing demand and the best that computing hardware can provide is widening at the rate of 8x per year or 500x every 3 years. This trend is not sustainable.
We need a new, much more brain-like, type of computing systems that are several orders of magnitude more complexity and energy efficient (see "Green AI" \cite{schwartz2019green}). 
Based on this work, we believe practically useful online sensory processing units that consume only few mm\textsuperscript{2} of die area and few mW of power are quite feasible in the near future. 
%


\bibliographystyle{IEEEtranS}
\bibliography{main_v3}

\end{document}